%
\magnification=\magstep1
\baselineskip = 0.75 true cm plus 0.04 true cm
\outer\def\beginsec#1\par{\filbreak\bigskip
  \message{#1}\centerline{\bf#1}\nobreak\smallskip\vskip-\parskip\noindent}
\hsize=15 true cm
\vsize=22 true cm
\def\today{\ifcase\month\or
  January\or February\or March\or April\or May\or June\or July\or
  August\or September\or October\or November\or December\fi
  \space\number\day, \number\year}
%
\footline={\tenrm\hss --~\folio~--\hss}
%

%
\def\hbn{\hfil\break\noindent}

\def\bn{\break\noindent}
%
\def\gsim{\buildrel >\over\sim }
\def\lsim{\buildrel <\over\sim }
\def\prl{Phys. Rev. Lett.}
\def\prb{Phys. Rev. B}

%

%
%
%
%
\countdef\ncit=101
\ncit=1
%
%
%
\edef\RFAA{$^{\the\ncit}$} \advance\ncit by 1 
\edef\RFBBa{$^{\the\ncit}$} \advance\ncit by 1 
\edef\RFBBb{$^{\the\ncit}$} \advance\ncit by 1 
\edef\RFBBc{$^{\the\ncit}$} \advance\ncit by 1 
\edef\RFBBd{$^{\the\ncit}$} \advance\ncit by 1 
\edef\RFCCa{$^{\the\ncit}$} \advance\ncit by 1 
\edef\RFCCb{$^{\the\ncit}$} \advance\ncit by 1 
\edef\RFCCc{$^{\the\ncit}$} \advance\ncit by 1 
\edef\RFCCd{$^{\the\ncit}$} \advance\ncit by 1
\edef\RFCCe{$^{\the\ncit}$} \advance\ncit by 1
\edef\RFDD{$^{\the\ncit}$} \advance\ncit by 1
\edef\RFEE{$^{\the\ncit}$} \advance\ncit by 1 
\edef\RFEEu{$^{\the\ncit}$} \advance\ncit by 1 
\edef\RFEEv{$^{\the\ncit}$} \advance\ncit by 1 
\edef\RFEEw{$^{\the\ncit}$} \advance\ncit by 1 
\edef\RFEEa{$^{\the\ncit}$} \advance\ncit by 1 
\edef\RFEExa{$^{\the\ncit}$} \advance\ncit by 1 
\edef\RFEEya{$^{\the\ncit}$} \advance\ncit by 1 
\edef\RFEEb{$^{\the\ncit}$} \advance\ncit by 1 
\edef\RFFF{$^{\the\ncit}$} \advance\ncit by 1 
\edef\RFFFxa{$^{\the\ncit}$} \advance\ncit by 1 
\edef\RFFFya{$^{\the\ncit}$} \advance\ncit by 1
\edef\RFFFza{$^{\the\ncit}$} \advance\ncit by 1 
\edef\RFFFa{$^{\the\ncit}$} \advance\ncit by 1 
\edef\RFFFb{$^{\the\ncit}$} \advance\ncit by 1 
\edef\RFFFc{$^{\the\ncit}$} \advance\ncit by 1 
\edef\RFFFd{$^{\the\ncit}$} \advance\ncit by 1 
\edef\RFFFg{$^{\the\ncit}$} \advance\ncit by 1 
\edef\RFFFh{$^{\the\ncit}$} \advance\ncit by 1 
\edef\RFGGa{$^{\the\ncit}$} \advance\ncit by 1 
\edef\RFGGb{$^{\the\ncit}$} \advance\ncit by 1 
\edef\RFHH{$^{\the\ncit}$} \advance\ncit by 1 
\edef\RFHHa{$^{\the\ncit}$} \advance\ncit by 1 
\edef\RFHHb{$^{\the\ncit}$} \advance\ncit by 1 
\edef\RFHHxb{$^{\the\ncit}$} \advance\ncit by 1 
\edef\RFHHc{$^{\the\ncit}$} \advance\ncit by 1 
\edef\RFHHd{$^{\the\ncit}$} \advance\ncit by 1 
\edef\RFHHub{$^{\the\ncit}$} \advance\ncit by 1 
\edef\RFIIa{$^{\the\ncit}$} \advance\ncit by 1 
\edef\RFIIb{$^{\the\ncit}$} \advance\ncit by 1 
\edef\RFIIc{$^{\the\ncit}$} \advance\ncit by 1 
\edef\RFIId{$^{\the\ncit}$} \advance\ncit by 1 
\edef\RFIIe{$^{\the\ncit}$} \advance\ncit by 1 
\edef\RFJJ{$^{\the\ncit}$} \advance\ncit by 1 
\edef\RFSFVertCorr{$^{\the\ncit}$} \advance\ncit by 1 
\edef\RFKK{$^{\the\ncit}$} \advance\ncit by 1 
\edef\RFIsoExpZ{$^{\the\ncit}$} \advance\ncit by 1 
\edef\RFIsoCritZ{$^{\the\ncit}$} \advance\ncit by 1 
\edef\RFIsoExpTcA{$^{\the\ncit}$} \advance\ncit by 1 
\edef\RFIsoExpTcB{$^{\the\ncit}$} \advance\ncit by 1 
\edef\RFPaScIsoD{$^{\the\ncit}$} \advance\ncit by 1 
\edef\RFAlleTr{$^{\the\ncit}$} \advance\ncit by 1 
\edef\RFDruOpt{$^{\the\ncit}$} \advance\ncit by 1 
\edef\RFEmBP{$^{\the\ncit}$} \advance\ncit by 1 
\edef\RFLL{$^{\the\ncit}$} \advance\ncit by 1
\edef\RFMM{$^{\the\ncit}$} \advance\ncit by 1 
\edef\RFNN{$^{\the\ncit}$} \advance\ncit by 1 
\edef\RFOO{$^{\the\ncit}$} \advance\ncit by 1 
\edef\RFPP{$^{\the\ncit}$} \advance\ncit by 1
%
%
%
\ncit=1
%
%
\edef\olRFAA{${\the\ncit}$} \advance\ncit by 1 
\edef\olRFBBa{${\the\ncit}$} \advance\ncit by 1 
\edef\olRFBBb{${\the\ncit}$} \advance\ncit by 1 
\edef\olRFBBc{${\the\ncit}$} \advance\ncit by 1 
\edef\olRFBBd{${\the\ncit}$} \advance\ncit by 1 
\edef\olRFCCa{${\the\ncit}$} \advance\ncit by 1 
\edef\olRFCCb{${\the\ncit}$} \advance\ncit by 1 
\edef\olRFCCc{${\the\ncit}$} \advance\ncit by 1 
\edef\olRFCCd{${\the\ncit}$} \advance\ncit by 1
\edef\olRFCCe{${\the\ncit}$} \advance\ncit by 1
\edef\olRFDD{${\the\ncit}$} \advance\ncit by 1
\edef\olRFEE{${\the\ncit}$} \advance\ncit by 1 
\edef\olRFEEu{${\the\ncit}$} \advance\ncit by 1 
\edef\olRFEEv{${\the\ncit}$} \advance\ncit by 1 
\edef\olRFEEw{${\the\ncit}$} \advance\ncit by 1 
\edef\olRFEEa{${\the\ncit}$} \advance\ncit by 1 
\edef\olRFEExa{${\the\ncit}$} \advance\ncit by 1 
\edef\olRFEEya{${\the\ncit}$} \advance\ncit by 1 
\edef\olRFEEb{${\the\ncit}$} \advance\ncit by 1 
\edef\olRFFF{${\the\ncit}$} \advance\ncit by 1 
\edef\olRFFFxa{${\the\ncit}$} \advance\ncit by 1 
\edef\olRFFFya{${\the\ncit}$} \advance\ncit by 1
\edef\olRFFFza{${\the\ncit}$} \advance\ncit by 1 
\edef\olRFFFa{${\the\ncit}$} \advance\ncit by 1 
\edef\olRFFFb{${\the\ncit}$} \advance\ncit by 1 
\edef\olRFFFc{${\the\ncit}$} \advance\ncit by 1 
\edef\olRFFFd{${\the\ncit}$} \advance\ncit by 1 
\edef\olRFFFg{${\the\ncit}$} \advance\ncit by 1 
\edef\olRFFFh{${\the\ncit}$} \advance\ncit by 1 
\edef\olRFGGa{${\the\ncit}$} \advance\ncit by 1 
\edef\olRFGGb{${\the\ncit}$} \advance\ncit by 1 
\edef\olRFHH{${\the\ncit}$} \advance\ncit by 1 
\edef\olRFHHa{${\the\ncit}$} \advance\ncit by 1 
\edef\olRFHHb{${\the\ncit}$} \advance\ncit by 1 
\edef\olRFHHxb{${\the\ncit}$} \advance\ncit by 1 
\edef\olRFHHc{${\the\ncit}$} \advance\ncit by 1 
\edef\olRFHHd{${\the\ncit}$} \advance\ncit by 1 
\edef\olRFHHub{${\the\ncit}$} \advance\ncit by 1 
\edef\olRFIIa{${\the\ncit}$} \advance\ncit by 1 
\edef\olRFIIb{${\the\ncit}$} \advance\ncit by 1 
\edef\olRFIIc{${\the\ncit}$} \advance\ncit by 1 
\edef\olRFIId{${\the\ncit}$} \advance\ncit by 1 
\edef\olRFIIe{${\the\ncit}$} \advance\ncit by 1 
\edef\olRFJJ{${\the\ncit}$} \advance\ncit by 1 
\edef\olRFSFVertCorr{${\the\ncit}$} \advance\ncit by 1 
\edef\olRFKK{${\the\ncit}$} \advance\ncit by 1 
\edef\olRFIsoExpZ{${\the\ncit}$} \advance\ncit by 1 
\edef\olRFIsoCritZ{${\the\ncit}$} \advance\ncit by 1 
\edef\olRFIsoExpTcA{${\the\ncit}$} \advance\ncit by 1 
\edef\olRFIsoExpTcB{${\the\ncit}$} \advance\ncit by 1 
\edef\olRFPaScIsoD{${\the\ncit}$} \advance\ncit by 1 
\edef\olRFAlleTr{${\the\ncit}$} \advance\ncit by 1 
\edef\olRFDruOpt{${\the\ncit}$} \advance\ncit by 1 
\edef\olRFEmBP{${\the\ncit}$} \advance\ncit by 1 
\edef\olRFLL{${\the\ncit}$} \advance\ncit by 1
\edef\olRFMM{${\the\ncit}$} \advance\ncit by 1 
\edef\olRFNN{${\the\ncit}$} \advance\ncit by 1 
\edef\olRFOO{${\the\ncit}$} \advance\ncit by 1 
\edef\olRFPP{${\the\ncit}$} \advance\ncit by 1
\centerline{\bf{Quasi-Particle dynamics of a}}
\centerline{\bf{strongly correlated polaron metal}}
\centerline{J. Zhong and H.-B. Sch\"uttler}
\centerline{Center for Simulational Physics}
\centerline{Department of Physics and Astronomy}
\centerline{University of Georgia}
\centerline{Athens, Georgia 30602}
\hbn
\centerline{Abstract}
We develop a simple diagrammatic 
low-energy boson exchange strong-coupling
model for an antiferromagnetically correlated polaronic metal.
The model is based on a simplified, local anharmonic
oscillator representation of the highly anharmonic
Born-Oppenheimer lattice potential which typically 
arises in a finite density polaron system, as a result
of polaronic self-localization. 
This Born-Oppenheimer potential exhibits 
a large manifold of nearly degenrate local
potential minima, and hence a new class of low-energy 
excitations, associated with tunneling processes
between the multiple Born-Oppenheimer wells.
In the present paper, we employ a simple Migdal-type 
single-exchange diagrammatic approximation in order 
to study the what happens when conduction electrons
are coupled to these anharmonic tunneling
excitations via conventional electron-phonon
interactions. Important low-temperature
features of such a model are 
(i) a  large polaronic mass enhancement $Z$ 
and a strongly narrowed, low-energy, 
near-Fermi-level van Hove resonance peak in the interacting
density of states at temperatures $T$ well below
the lattice tunneling energy scale $\Omega_t$;
(ii) a rapid thermal suppression of $Z$ and of the van Hove resonance 
which occurs when the temperature $T$ becomes comparable to the
lattice tunneling excitation energy scale $\Omega_t$;
(iii) a broad incoherent background in the single-particle spectra
$|G^{\prime\prime}({\vec{k}},\omega)|$ near the Fermi energy;
(iv) strongly flattened quasi-particle bands
over wide regions of the Brillouin zone near the van Hove points;
and
(v) strong isotopic mass dependence of the electronic
mass enhancement $Z$ at temperatures $T$ well below $\Omega_t$
which, along with $Z$, becomes rapidly suppressed for $T$ 
comparable to $\Omega_t$.

Surprisingly, in spite of the large low-$T$
mass enhancment, the anharmonic lattice fluctuations
in the high temperature regime $T\gsim\Omega_t$,
give rise only to a moderate, essentially $T$-independent
quasi-particle damping. This damping does {\it not} suppress coherent 
quasi-particle propagation in a wide, physically relevant 
parameter regime. We thus argue that, in a ``weakly bound''
finite-density polaron system, a novel polaron scenario may be realized
where only the lattice motion, but not the quasi-particle
propagation become incoherent in the high-$T$ regime.
This scenario is in contrast to the conventional small-polaron
strong-coupling picture, where the quasi-particle transport
inevitably becomes incoherent at temperatures exceeding the
polaron tunneling energy scale. 

We also discuss possible implications of the anharmonic
tunneling model for the cuprate superconductors.
The predicted isotopic mass dependence of the low-$T$ ($T<\Omega_t$) 
electron quasi-particle mass enhancement provides a unique
signature for the involvemnt of anharmonic tunneling
fluctuations. We suggest possible low temperature 
isotope experiments to explore this question in the cuprates.  
\hbn
\hbn
PACS 74.20.-z., 75.10.-b., 63.20.Kr, 71.30.+h, 71.45.Lr
\hbn
\vfil
\eject
\centerline{\bf I. INTRODUCTION}

Hubbard-type strongly correlated electron
models{}{}{}{\RFAA} can adequately describe the 
low-lying charge and anti-ferromagnetic (AF) spin
excitations in the undoped, insulating cuprate systems.{}{}{}{\RFBBa$^{-}$\RFBBd} 
However, in the doped systems, the low-energy physics
may be substantially affected by coupling to the
lattice degrees of freedom,
since, {\it{in the presence of AF spin correlations}},
already a rather weak electron-phonon coupling
can cause polaronic self-localization of the dopant-induced 
charge carriers.{}{}{}{\RFCCa$^{-}$\RFCCe}
As a consequence, large-amplitude
anharmonic lattice displacement fluctuations occur, resulting
from polaron tunneling motion.{}{}{}{\RFCCa$^{-}$\RFCCb} 
In the cuprate high-$T_c$ systems, there is indeed evidence
for polaron formation and for large-amplitude, anharmonic
lattice tunneling fluctuations of possible polaronic origin.{}{}{}{\RFDD}
However, the observed quasi-particle mass enhancements $Z\!\sim\!2-4$
in the normal state of the doped, metallic cuprates{}{}{}{\RFEE$^{-}$\RFEEb} 
fall far short of the magnitudes $Z\!\gsim\!10-10^3$ (or larger)
which one would naively expect to observe in a polaronic system.

In the present paper, we propose a new modeling approach towards
the normal-state electron quasi-particle dynamics of an antiferromagnetically 
correlated polaronic metal where the essential features
of the polaronic lattice
\bn
dynamics{}{}{}{\RFCCa$^{-}$\RFCCb} and AF spin fluctuation
exchange{}{}{}{\RFFF$^{-}$\RFFFh} are combined into
a generalized Migdal-type diagrammatic treatment.
The polaronic lattice anharmonicity does then
indeed lead to  large polaronic mass enhancement $Z$ 
when the temperature $T$ {\it{and}} excitation energies $\omega$
are well below a certain, generally small lattice
tunneling energy scale $\Omega_t$. 

However, for $T\!\gsim\!\Omega_t$,
the polaronic contribution to $Z$ is suppressed 
and $Z$ is dominated by the smaller spin fluctuation contribution.
The polaronic lattice fluctuations contribute then only
to the quasi-particle damping, in a manner quite similar to
-- and experimentally indistinguishable from -- disorder
scattering. Surprisingly, in spite of the large
low-$T$ mass enhancement, the high-$T$ quasi-particle
damping caused by the anharmonic lattice fluctuations
is quite modest and the electronic transport dynamics retains
essentially coherently propagative character.
This suggests the possibility of a heretofore
unexplored ``weakly bound'' polaron scenario
wherein polaron formation does {\it not}
inevitably lead to complete incoherence at temperatures
exceeding the polaron tunneling energy scale.
Our treatment thus suggests
a novel approach towards {\it{dense}} polaron systems
which bridges the gap between the conventional
Fermi-liquid-based Migdal theory of delocalized, harmonic
electron-phonon systems{}{}{}{\RFGGa$^{,}$\RFGGb} 
and the strong-coupling methods
developed originally for the strongly localized, single-electron
small-polaron problem.{}{}{}{\RFHH$^{-}$\RFHHd}

The remainder of the paper is organized as follows:
In Section II, we outline the basic underlying ideas of polaronic
anharmonicity in the lattice dynamics of a strongly correlated
electron-phonon system; we propose a simple exactly numerically 
solvable model for the polaronic Born-Oppenheimer potential;
and we discuss the effective electron-electron interaction $V_{AP}$
which is mediated by exchange of the anharmonic tunneling
excitations arising from the lattice quantum dynamics within such 
an ``anharmonic phonon'' potential.
We then combine this lattice-mediated electron
interaction $V_{AP}$ with a simple
electron-electron model interaction $V_{SF}$ describing
AF spin fluctuation exchange. The full model interaction
$V=V_{AP}+V_{SF}$ then serves as our input into a 
self-consistent diagrammatic single-electron self-energy calculation,
which we formulate at the level of a 
Migdal-type single-exchange approximation
for a single-band conduction electron system on a two-dimensional (2D)
square lattice.
From the self-energy, we extract important physical quantities
such as the electron quasi-particle dispersion $E_{\vec{k}}$, and
and the quasi-particle mass enhancement $Z$, the single-electron
spectral function $G^{\prime\prime}(\vec{k},\omega)$ and the 
the momentum-integrated density of states $\rho(\omega)$.

In Section III, we discuss basic physical principles 
governing the polaronic mass enhancement in our
diagrammatic model. We show, by analytical 
arguments, how, within our diagrammatic model, one can recover,
at least qualitatively, the central results for polaronic
mass enhancement which are conventionally derived within
a non-perturbative electron-phonon strong-coupling
theory. Important implications for a potential observable
isotopic mass dependence in the electron mass enhancement
are discussed in the anharmonic tunneling exchange model
and contrasted with a corresponding harmonic phonon
exchange model. Based on self-consistent numerical
self-energy calculations, we then discuss the 
temperature ($T$-) dependence of
the polaronic self-energy and its implications for 
the $T$-dependence of the quasi-particle mass enhancement
$Z$ and of the quasi-particle damping.
We then consider the effects of anharmonic tunneling
fluctuation exchange on the momentum integrated density of states
and, in particular on the van Hove singularity.
Numerical results for the single-particle spectral functions
and for the overall quasi-particle band dispersion
in the full Brillouin zone will also be discussed.

Lastly, in Section IV, we present a summary of our results,
with emphasis on potential experimental implications
of our model calculations for the cuprate high $T_c$
materials. We conclude
with an outline of the generic features of our model
which are likely to be of general physical significance
and likely to be recovered in a more realistic polaron model.
\hbn
\hbn
\hbn
\centerline{\bf 
II. LATTICE TUNNELING FLUCTUATION EXCHANGE MODEL}

\hbn
\centerline{\bf A. Polaronic anharmonicity}

As a simplest model, we start from a two-dimensional (2D) 
antiferromagnetically correlated Hubbard-type
single-band electron system on a 2D square 
lattice,{}{}{}{\RFAA$^{-}$\RFBBd$^{,}$\RFFF$^{-}$\RFFFh} coupled to
an Einstein optical phonon system 
via a Holstein electron-phonon interaction{}{}{}{\RFCCa$^{-}$\RFCCe$^{,}$\RFHH$^{-}$\RFHHd}
$$
H_{EP}=\sum_j C u_j(n_j-1).
\eqno{(1)}
$$
Here, $C$ denotes the deformation potential coupling constant,
$n_j\!=\!0,1,2$ the electron occupation, and $u_j$
the local Einstein oscillator displacement at 
lattice sites $j\!=\!1...N$. 
Physically, $u_j$ could represent, for example, 
a collective coordinate for local breathing-type displacements
of planar or apical oxygens in the cuprates.

In the absence of EP coupling,
the lattice dynamics is governed by a {\it bare} harmonic
restoring potential $W_P(\{u_j\})$ 
corresponding to a non-interacting phonon
Hamiltonian
$$
H_P = \sum_j{1\over 2M} p_j^2 + W_P(\{u_j\})
\eqno{(2)}
$$
where $M$ is the atomic or molecular mass associated
with each displacement degree of freedom $u_j$
and $p_j\equiv -i\hbar\partial/\partial u_j$ is the
conjugate momentum of $u_j$. 
In the simplest microscopic strongly correlated
electron-phonon model, the Holstein-Hubbard 
model{}{}{}{\RFCCa$^{,}$\RFCCb$^{,}$\RFCCd$^{,}$\RFHH$^{-}$\RFHHd} 
$W_P$ is given by an Einstein potential
$$
W_P(\{u_j\}) = \sum_j {1\over 2} K u_j^2
\eqno{(3)}
$$
with a bare, harmonic restoring force constant $K$.
To be specific, we will, for the following
discussion, adopt the Holstein-Hubbard
model as our underlying microscopic model Hamiltonian.
The basic ideas are more general, however,
and can, in principle, be extended 
to more complicated electron-phonon
systems exhibiting polaronic self-localization.
The EP coupling strength can then be quantified 
by a characteristic energy 
$$
E_P \ =\ {C^2\over K}
\eqno{(4)}
$$
which measures the strength of the phonon-mediated
on-site attraction, as well as the ingle-polaron
binding energy in the ionic (zero-bandwidth limit)
of the Holstein-Hubbard model.

In the nearly $1\over2$-filled Holstein-Hubbard electron
system with sufficient EP coupling strength $E_P$, 
a dopant-induced carrier of, e.~g., hole-type, $n_j\!=\!0$,
self-localizes.{}{}{}{\RFCCa$^{-}$\RFCCd}
That is, by locally distorting the lattice by some 
$\Delta u_j\!\equiv\!d_t$,
the carrier creates an attractive electron potential well, of depth
$$
\Delta_P \equiv C d_t
\eqno{(5)}
$$, 
and thus lowers its energy by forming 
a bound state.{}{}{}{\RFCCa$^{-}$\RFCCd$^{,}$\RFHH$^{-}$\RFHHd}
The inter-site tunneling motion of such polaronic carriers
causes the displacement $u_j$ at each site $j$ to fluctuate between two
distinct equilibrium positions, $u_\pm$, with $|u_+\!-\!u_-|\!\cong\!d_t$.

As explained at length in 
Refs.~[\olRFCCa,\olRFCCb,\olRFHHub],
this physical picture can be fully captured 
within the framework of the Born-Oppenheimer approximation, 
that is, in terms of a renormalized
lattice potential $W(\{u_j\})$.{}{}{}{\RFCCa$^{,}$\RFCCb$^{,}$\RFHHub}
In the full microscopic treatment, based 
on the Holstein-Hubbard model,{}{}{}{\RFCCa$^{,}$\RFCCb} this Born-Oppenheimer
potential is given by the total ground state energy
of the 0th order adiabatic Hamiltonian $H_{ad}(\{u_j\})$
which depends parametrically on the fixed, $c$-number 
lattice displacements $\{u_j\}$. 
$H_{ad}(\{u_j\})$ consists of the bare potential $W_P(\{u_j\})$, the 
EP coupling term $H_{E}(\{u_j\})$ and the 
purely electronic part $H_E$ which is independent
of the displacements $u_j$. Tn the case of the Holstein-Hubbard
model, $H_E$ is just the standard Hubbard Hamiltonian.{}{}{}{\RFAA$^{,}$\RFBBa$^{-}$\RFBBd}
Detailed microscopic studies along these lines, based on analytical
considerations and on numerical many-body total energy calculations
on finite clusters have been carried out.{}{}{}{\RFCCa$^{,}$\RFCCb$^{,}$\RFCCd}

These microscopic studies reveal that the transition or cross-over
from the delocalized carrier regime to the self-localized carrier
regime is fundamentally accompanied by a qualitative change
in the character of the renormalized lattice potential $W$.
In the delocalized carrier regime, the low-energy lattice
dynamics is governed essentially by one lowest, absolute
potential minimum in $W(\{u_j\})$. 
Quadratic expansion of $W(\{u_j\})$ around its minimum,
thus leads again to an essentially 
harmonic lattice dynamics where the primary 
renormalization effect of the EP coupling
is to modify the harmonic restoring force constants, 
relative to $W_P(\{u_j\})$. 

In the self-localized carrier regime, on the other hand,
$W(\{u_j\})$ acquires a highly degenerate, 
anharmonic, multiple-minimum low-energy structure
wherein different local minima of $W$ correspond to different
possible real-space configurations of the self-localized carriers.
The polaron tunneling dynamics in this Born-Oppenheimer 
picture can then be described in terms of lattice tunneling 
processes between ``adjacent''$W$-minima.
The low-energy structure of the Born-Oppenheimer energy surface
is thus highly complex, especially, at a finite
polaron density, involving, in general highly
anharmonic, highly non-local multi-site interactions
betwen the $u_j$ coordinates. The low-energy lattice
tunneling dynamics within this degenrate multi-well structure
is further complicated by the presence of non-trivial
$e^{i\pi}$ Berry phase factors which arise from 
adiabatic electron groundstate wavefunction
overlaps during the adiabatic tunneling
motion of the lattice between different local $W$-wells.{}{}{}{\RFCCb}
Trying to calculate, or even simulate, the low-energy
lattice dynamics in the self-localized regime,
from a ``realistic'', that is, microscopically based
Born-Oppenheimer potential does not appear to be feasible
at the pesent time, especially in the physically
most interesting case of finite polaron carrier densities.

We are therefore proposing to
simplify the polaronic lattice dynamics problem 
by replacing the microscopic lattice 
potential with a drastically simplified model potential.
This simplified potential does not, by any measure, capture 
all the intricate details of the microscopic potential.
Nevertheless, as we will show, it retains some of the 
essential physical features of the polaronic lattice dynamics.
Furthermore, this model potential can serve as the starting point
for more realistic modeling approaches which may capture
some of those physical aspects, such as non-local multi-site
anharmonicity, which our present model fails to include.

Specifically, we propose to {\it{model}} 
the polaronic lattice potential by the following ansatz
$$
W(\{u_j\}) = \sum_j w(u_j) \equiv \sum_j 8\Delta_B[2(u_j/d_t)^4-(u_j/d_t)^2)]
\ .
\eqno{(6)}
$$ 
The single-site potential
$
w(u_j)
$
is a local (on-site) double-well
with  minima at 
$$
u_{\pm}\!=\!\pm d_t/2\ .
\eqno{(7)}
$$
The two potential minima in $w(u_j)$ are 
separated by a tunneling barrier of height $\Delta_B$ and width $d_t$.

Eq.~(6) captures some of the essential physics of the polaronic
lattice dynamics, namely, the fact that each $u_j$ undergoes large-amplitude 
fluctuations between two distinct equilibrium positions $u_{\pm}$.
Neglected in this simple {\it{model}} potential
are are all the anharmonic inter-site correlations described
above within the microscopic framework.{}{}{}{\RFCCa$^{,}$\RFCCb}

The essential new feature of the polaronic lattice dynamics,
which is absent in harmonic phonon systems,
is a new class of large-amplitude,
low-energy inter-well tunneling excitations, 
which exist in addition to the 
familiar, small-amplitude, intra-well phonon-like
excitations. In our model, Eq.~(6), these two types of 
excitations are represented by the excitation
energies $\Omega_t$ and $\Omega_h^\prime$ of, respectively, the 1st and 2nd 
excited state, measured from the groundstate, in the single-site
double-well $w(u)$. For reasonable parameters, we find 
$$
\Omega_t \ll \Omega_h^\prime
\eqno{(8)}
$$ 
where $\Omega_h^\prime$ is comparable to
the single-well harmonic phonon energy
which can be estimated by
$$
\Omega_{h} \equiv \hbar\Big({K_h\over M}\Big)^{1/2}
\eqno{(9)}
$$
where the single-well hamonic restoring force
$$
K_h \equiv {\partial^2 \over\partial u^2}  w(u_{\pm}) 
=40 {\Delta_B \over d_t^2}
\eqno{(10)}
$$ 
is determined by the curvature of the potential
at the single-well minimum and $M$ is the atomic/molecular
mass associated with the local oscillator degree of freedom $u$.
In a full microscopic Holstein-Hubbard calculation,{}{}{}{\RFCCa}
$\Omega_{h}$ is generally comparable to the bare ($C\!\equiv\!0$) harmonic
phonon energy $\Omega_{h0}=\hbar(K/M)^{1/2}$.

\hbn
\centerline{\bf B. Fluctuation exchange potentials}

By exchange of anharmonic lattice excitations
an effective electron-electron interaction
$V_{AP}$ is mediated.{}{}{}{\RFIIa$^{-}$\RFIIe} As in standard
{\it{harmonic}} phonon exchange models,{}{}{}{\RFGGa$^{,}$\RFGGb}
this anharmonic phonon exchange potential $V_{AP}(\vec{q},i\omega)$ 
can be expressed as a product
of electron-lattice coupling constants [i.e $C^2$ in our model, Eq.~(1)]
and appropriate lattice displacement correlation functions.{}{}{}{\RFIIa$^{-}$\RFIIe}
That is, in our model, following Refs.~[\olRFIIa-\olRFIIe],
$$
V_{AP}(\vec{q},i\omega)
=-C^2 \int_0^\beta d\tau {1\over N} \sum_{j,\ell}
e^{i\omega\tau-i\vec{q} \cdot (\vec{r}_j-\vec{r}_\ell)}
 \langle u_j(\tau)_{AP} u_\ell(0)\rangle_{AP} \ .
\eqno{(11)}
$$
for an electron-electron momentum-energy transfer $(\vec{q},i\omega)$
where $i\omega\equiv 2n_\omega\pi T$ is the even Matsubara frequency
with integer $n_\omega$.
$N$ denotes the number of lattice sites on an $N=L\times L$
2D square lattice lattice with periodic boundary conditions, 
$\vec{r}_j$ and $\vec{r}_\ell$ 
are the position vectors of lattice sites $j$
and $\ell$, respectively, $\beta\equiv 1/T$ is the inverse
temperature. The notations $\langle ... \rangle_{AP}$ 
and  $u_j(\tau)_{AP}$ in Eq.{(11)}
indicate thermal averaging and, respectively, evolution in
imaginary time $\tau$, according to a Born-Oppenheimer
effective lattice Hamiltonian given by
$$
H_{AP} 
= \sum_j {1\over 2M} p_j^2 + W(\{u_j\})
\equiv \sum_j h_{u_j}\ .
\eqno{(12)}
$$
For our simple local 
model potential, Eq.~(6), $H_{AP}$ 
describes $N$ uncoupled local double-well oscillator
degrees of freedom $u_j$, each subject to the local Hamiltonian
$$
h_u = {1\over 2M} p^2 + w(u) \ ,
\eqno{(13)}
$$
which acts only on a single local degree of freedom $u$.

Because the simple on-site coupling, Eq.~(1), and local
model potential,
\bn
Eq.~(6), $V_{AP}\equiv V_{AP}(i\omega)$ 
becomes $\vec{q}$-independent in our model,
and can be expressed in terms of the
local (single-site) eigenstates of $h_u$.{}{}{}{\RFIIa$^{-}$\RFIIe}
Namely, written in terms of its spectral representation,
$$
V_{AP}(i\omega) = \int_{-\infty}^{+\infty} {d\omega^\prime \over \pi}
{V_{AP}^{\prime\prime}(\omega^\prime) \over \omega^\prime - i\omega}
\eqno{(14)}
$$
where the spectral function on the real-$\omega^\prime$ axis,
$V_{AP}^{\prime\prime}(\omega^\prime)\equiv
{1\over2} [V_{AP}(\omega^\prime+i0^+)-V_{AP}(\omega^\prime-i0^+)]$,
is given by
$$
V_{AP}^{\prime\prime}(\omega^\prime) = 
{-\pi} \sum_{\iota,\kappa}
|\langle \psi_\kappa | u | \psi_\iota \rangle |^2
{e^{-\beta \eta_\iota} \over z_h}
[\delta(\eta_\kappa - \eta_\iota - \omega^\prime) -
\delta(\eta_\kappa - \eta_\iota + \omega^\prime)] \ .
\eqno{(15)}
$$
Here, $\psi_\iota(u)$ denotes the normalized eigenstates
of $h_u$ with eigenenergy $\eta_\iota$, labeled in ascending
order, beginning with the groundstate $\iota\equiv 0$.
Also, $z_h\equiv\sum_\iota exp(-\beta \eta_\iota)$ 
is the corresponding single-site partition function.
$V_{AP}^{\prime\prime}$ can thus be easily evaluated
by solving numerically the single-site Schr\"odinger 
equation for $h_u$, to obtain $\psi_\iota$ and $\eta_\iota$.
By carrying out the $\iota$- and $\kappa$- summation in Eq.~(15)
numrically up to, say, the 5--th excited state, $\iota,\kappa\le 5$,
one captures over $99\%$ of the total spectral
weight in $V_{AP}^{\prime\prime}$, assuming physical parameter values
given below.

We augment $V_{AP}$ by a simple model for
the electron-electron interaction potential,
$V_{SF}(\vec{q},i\omega)$, 
mediated by the AF spin fluctuations.{}{}{}{\RFFF$^{-}$\RFFFh} 
The model potential is defined for the 
2D square lattice {\it via} its spectral spectral representation
$$
V_{SF}(\vec{q},i\omega) = \int_{-\infty}^{+\infty} {d\omega^\prime \over \pi}
{V_{SF}^{\prime\prime}(\vec{q},\omega^\prime) \over \omega^\prime - i\omega}
\eqno{(16)}
$$
with a spectral function{}{}{}{\RFJJ} of the form{}{}{}{\RFFFa}
$$
V_{SF}^{\prime\prime}({\vec{q}},\omega) = -g_s^2\sum_{\vec{Q}^*}
F(\vec{q}-\vec{Q}^*)\Phi^{\prime\prime}(\omega) \ .
\eqno{(17)}
$$
A simple continuous, $\vec{q}$-independent 
linear spin fluctuation spectrum is assumed, with{}{}{}{\RFFFa}
$$
\Phi^{\prime\prime}(\omega) = {2\omega \over \Omega_s^2} 
\Theta(\Omega_s - |\omega|) \ .
\eqno{(18)}
$$
extending up to a typical spin fluctuation cutoff $\Omega_s$.
The $\vec{q}$-dependence is assumed to be Lorentzian{}{}{}{\RFFFa}
$$
F(\vec{q}-\vec{Q}^*)= 
{A_{SF} \over |\vec{q}-\vec{Q}^*|^2 + \kappa_{AF}^2 }
\eqno{(19)}
$$ 
with a half-width $\kappa_{AF}$ of order of 
the inverse AF correlation length.
The $\vec{Q}^*$ summation in Eq.~(17) is
over the four equivalent AF wavevectors
$$
\vec{Q}^{*} =(\pi,\pi),\ (-\pi,\pi),
\ (-\pi,\pi),\ (-\pi,-\pi) \ .
\eqno{(20)}
$$
The overall normalization factor $A_{SF}$ is chosen so that
$$
g_s^2 =
 {-1\over N}\sum_{\vec{q}}
 \int_0^\infty d\omega V_{SF}^{\prime\prime}({\vec{q}},\omega) 
\eqno{(21)}
$$
where $g_s^2$ is the spin fluctuation
coupling constant and the $\vec{q}$-summation is only 
over the 1st Brillouin zone of the 
2D square lattice with lattice constant $a$.
Also, Eq.~(17) is understood to apply only 
to $\vec{q}$-points within the
1st Brillouin zone, with periodic continuation
beyond the 1st zone boundaries.

We should emphasize here again that the foregoing
simplified model potentials can
reproduce only the most qualitative physical features
of the order of magnitude, of the $T$-variation
and of the $\omega$-dependences of various physical
quantities discussed below. For instance,
no $T$-dependence is built into the spin fluctuation
exchange potential. Likewise, the spectral and thermal
properties of the polaronic lattice dynamics
has been severely simplified here, to retain physical
features such as large-amplitude tunneling fluctuations
and a low tunneling ($\Omega_t$) energy scale,
while neglecting, {\it e.~g.}, all effects from inter-site lattice
couplings which, surely, will alter the details of the
$T$- and $\omega$-dependence of $V_{AP}^{\prime\prime}$.
Therefore, these potentials cannot and should not be
expected to reproduce, say,  details of the 
quasi-particle spectral shapes or of the $T$-dependence
of quasi-particle lifetimes.

\hbn
\centerline{\bf C. Self-energy calculation}

From the total interaction $V\!\equiv\!V_{AP}+V_{SF}$,
we obtain the single-electron self-energy $\Sigma$
in a self-consistent Migdal single-exchange 
approximation.{}{}{}{\RFGGa$^{,}$\RFGGb$^{,}$\RFJJ}
In the Matsubara frequency domain, this single-exchange
diagram for the single-electron self-energy is given by{}{}{}{\RFGGa$^{,}$\RFGGb$^{,}$\RFJJ}
$$
\Sigma(\vec{k},i\nu) = 
-{T\over N} \sum_{\vec{k}^\prime,i\nu^\prime}
V(\vec{k}-\vec{k}^\prime,i\nu-i\nu^\prime) 
G(\vec{k}^\prime,i\nu^\prime)
\eqno{(22)}
$$
where $\vec{k}$ and $\vec{k}^\prime$ denote electron momenta
and $i\nu\equiv (2n_\nu+1)\pi T$ and 
$\nu^\prime (2n_\nu^\prime+1)\pi T$ are odd Matsubara
frequncies with integer $n_\nu$ and $n_\nu^\prime$.

The fully dressed electron Green's function in Eq.~(22) 
is given by the Dyson equation
$$
G(\vec{k},i\nu) = 
{1\over i\nu-\epsilon_{\vec{k}}-\Sigma(\vec{k},i\nu) }
\eqno{(23)}
$$
where $\epsilon_{\vec{k}}$ is the non-interacting
single-electron conduction bandstrucure. We are
adopting a single-band tight-binding model 
with 1st neighbor hybridization $t$ only, again
defined on the 2D square lattice appropriate to a
single $CuO_2$ layer in the cuprates. Hence,
for $\vec{k}\equiv(k_x,k_y)$,
$$
\epsilon_{\vec{k}} = -2t[\cos(a k_x) + \cos(a k_y)] -\mu
\eqno{(24)}
$$
with the chemical potential $\mu$ absorbed
in $\epsilon_{\vec{k}}$.

The basic idea underlying this diagrammatic approach is 
that, at the level of the Born-Oppenheimer approximation,
all renormalizations of the phonon system due to EP coupling
are, in principle, contained in the effective lattice potential
$W$. Hence, the displacement correlation function
in Eq.~(11) represents, in principle, the fully dressed
phonon propagator and $V_{AP}$ must, in this context,
be regarded as the already fully screened phonon-mediated 
electron-electron interaction potential, 
which can be directly inserted into the Migdal single-exchange
diagram, without further renormalization.
Beyond the well-controlled Born-Oppenheimer approximation,
there are thus two primary further approximations being made here:
firstly, the replacement of the quite complex polaronic
multiple-well lattice potential by the simple local double-well
model, Eq.~(6), and, secondly, the neglect of higher-order
vertex corrections, in the Migdal approximation.
Of these two, the former is likely to be the more severe
one; the latter can actually be roughly justified
for the model parameter values employed below,
based on standard order of magnitude estimates for the leading
order vertex correction. Another essential --
and likely uncontrolled -- approximation is of course 
the treatment of AF spin fluctuation exchange within the Migdal 
approximation.{}{}{}{\RFSFVertCorr} This constitutes a common,
fundamental problem in all presently
existing diagrammatic AF spin fluctuation
exchange theories. We have, in that regard, no further
progress to offer in the present paper.

By standard analytical continuation techniques,{}{}{}{\RFJJ}
Eq.~(23) can be recast into the real-frequency domain
to yield the self-energy spectral function 
$\Sigma^{\prime\prime}$
on the real-$\omega$ axis, expressed in terms of the
spectral functions of $V$ and $G$, namely
$$
\eqalignno{
\Sigma^{\prime\prime} (\vec{k},\omega)
&\equiv 
{1\over 2} 
[\Sigma(\vec{k},\omega+i0^+) - \Sigma(\vec{k},\omega-i0^+)]
\cr
&=
{1\over N}\sum_{\vec{k}^\prime}
\int_{-\infty}^{+\infty} {d\omega^\prime \over \pi}
V^{\prime\prime}(\vec{k}-\vec{k}^\prime,\omega-\omega^\prime)
G^{\prime\prime}(\vec{k}^\prime,\omega^\prime)
\cr
&
\ \ \ \ \ \ \ \ \ \ \ \ \ 
\ \ \ \ \ \ \ \ \ \times
[b(\omega-\omega^\prime)+1-f(\omega^\prime)] \ .
&{(25)}
\cr
}
$$
Here, $f(\omega)\equiv 1/(e^{\beta\omega}+1)$
and $b(\omega)\equiv 1/(e^{\beta\omega}-1)$
denote the Fermi and Bose factor, respectively,
and the $\vec{k}^\prime$-integral is over the
1st Brillouin zone of the 2D square lattice.
The spectrum of the single-electron Green's
function in Eq.~(25) is given by
$$
\eqalignno{
G^{\prime\prime}(\vec{k},\omega)
&\equiv
{1\over2}
[G({\vec{k}},\omega+i0^+) - G({\vec{k}},\omega-i0^+)]
\cr
&= 
{
\Sigma^{\prime\prime}(\vec{k},\omega)
\over
[\omega-\epsilon_{\vec{k}}-\Sigma^\prime(\vec{k},\omega)]^2
+[\Sigma^{\prime\prime}(\vec{k},\omega)+0^+]^2
}
&{(26)}
\cr
}
$$
and  $\Sigma^\prime$ is obtained from the self-energy 
spectral function $\Sigma^{\prime\prime}$,
via the Kramers-Kronig relation
$$ 
\Sigma^\prime(\vec{k},\omega)
\equiv
{1\over 2}
[\Sigma^\prime(\vec{k},\omega+i0^+)+
\Sigma^\prime(\vec{k},\omega-i0^+)]
=
\int\!\!\!\!\!\!P
{d\omega^\prime\over \pi}
{
\Sigma^{\prime\prime}(\vec{k},\omega^\prime)
\over 
\omega^\prime-\omega 
} 
\eqno{(27)}
$$
where $\int\!\!\!\!\!\!P$ denotes the principal value
integral.

Eqs.~(25)-(27) are solved iteratively, starting from
the 0--th order ansatz for the self-energy $\Sigma_0\equiv 0$.
To evaluate $\Sigma^{\prime\prime}$ for $N\!\to\!\infty$
numerically in the real-frequency domain, Eq.~(25), with
high $\omega$-resolution, we use a modified, 2D
adaptation of the so-called ``tetrahedron method''{}{}{}{\RFKK} 
to carry out the $\vec{k}^\prime$-integral in Eq.~(25).
Specifically, to handle the $\vec{k}^\prime$
integration over the nearly singular
quasi-particle peak structure of $G^{\prime\prime}$
efficiently, the {\it inverse} of the
electron Green's, from Eq.~(23), that is
$
1/G({\vec{k}},\omega\pm i0^+)
$
[{\it not} $G({\vec{k}},\omega\pm i0^+)$]
and, separately, 
the $\vec{q}$-dependent factor of the
interaction potential,
$V^{\prime\prime}$, from Eqs.~(17),(18),
are linearly interpolated over small triangular 
$\vec{k}$-elements. Upon, expressing 
$G^{\prime\prime}\equiv {\rm Im} G$ 
in terms of the $G$, so interpolated,
the $\vec{k}^\prime$-integral over each such triangular
element can be calculated analytically.
To construct the triangular interpolation grid, 
the Brillouin zone is covered with an $N=L\times L$ grid of 
equal square elements where each square is then subdivided
along a diagonal into two equal triangles.
To preserve the 4-fold point symmetry of the lattice
in the resulting triangular grid,
the subdividing diagonal is chosen to be along the 
$(+1,+1)$-direction for square elements in the 1st and 3rd quadrant
[{\it i.~e.} $\{\vec{k}=(k_x,k_y)\ |\ k_xk_y>0\}$],
and along the
$(+1,-1)$-direction for square elements in the 2nd and 4th quadrant
[{\it i.~e.} $\{\vec{k}=(k_x,k_y)\ |\ k_xk_y<0\}$]
of the Brillouin zone. The numerical results
shown below were obtained on grids
with $2N=2\times40\times40$ and  $2N=2\times80\times80$
triangular elements.
Comparisons between the two grid sizes at a few selected
parameter values, including $T=0$, indicate general agreement 
of the self-energy results at the $1-2\%$-level or better.

The chemical potential $\mu$ is adjusted so that
the electron concentration{}{}{}{\RFJJ}
$$
\langle n_j \rangle = 
{-2\over N}\sum_{\vec{k}} \int_{-\infty}^{+\infty}
f(\omega) G^{\prime\prime}(\vec{k},\omega)
\eqno{(28)}
$$
equals the specified input value. Note that the prefactor
$2$ arises from summation over the electron spin and 
$f(\omega)\equiv 1/(e^{\beta\omega}+1)$
denotes again the Fermi factor.

The self-consistent quasi-particle band $E_{\vec{k}}$
is then obtained from the solution of{}{}{}{\RFJJ}
$$
E_{\vec{k}}\!=\epsilon_{\vec{k}}+\Sigma^\prime(k,E_{\vec{k}}) \ .
\eqno{(29)}
$$
The interacting Fermi surface is defined as the locus
of all $\vec{k}$-points $\vec{k}_F$ where
$$
E_{\vec{k}_F} = 0 \ .
\eqno{(30)}
$$
The mass enhancement factor on the Fermi surface is{}{}{}{\RFJJ}
$$
 Z(\vec{k}_F)
  \cong 1-\partial_{\omega}\Sigma^\prime({\vec{k}}_F,\omega\!=\!0).
\eqno{(31)}
$$
Corrections to $Z$ due to{}{}{}{\RFJJ}
$
\partial_{\vec{k}}\Sigma^\prime
$
are $\lsim\!5-15\%$, for the parameters studied, 
and henceforth neglected.
\hbn
\hbn
\hbn
\centerline{\bf III. RESULTS AND DISCUSSION}

\hbn
\centerline{\bf A. Selfenergy and mass enhancement}

Before discussing the numerical results obtained with the
above described method, we will first try to obtain a qualitative
analytical understanding of the self-energy
and mass enhancement results
obtained at low excitation energies and temperatures 
$|\omega|, T \ll\Omega_t$.

For $T\ll\Omega_t$,
the mass enhancement $Z$ is dominated by the low-energy 
($\sim\!\Omega_t$) lattice tunneling excitations.
To estimate this, let us first consider the tunneling contribution
to the potential, $V_t$, in Eq.~(15).
By isolating the low-$T$ tunneling term, that is, the contribution
arising from the excitation $\iota\!=\!0\to\kappa=1$ from the
ground- to the 1st excited state, we get
$$
V_t^{\prime\prime}(\omega) = 
{-\pi}
|\langle \psi_1 | u | \psi_0 \rangle |^2
[\delta(\Omega_t - \omega) -
\delta(\Omega_t + \omega)] \ ,
\eqno{(32)}
$$
using $\Omega_t\equiv \eta_1-\eta_0$ and the fact
that the Boltzmann weight filters out the groundstate
in Eq.~(15) at $T\ll\Omega_t$, {\it i.~e.},
$z_h^{-1}\exp(-\beta\eta_\iota)\cong \delta_{\iota,0}$.

To estimate the transition matrix element in Eq.~(32)
we introduce the ``single-well'' basis states
$$
\psi_\pm(u)\ \equiv\ {1\over \sqrt{2} }\ 
                     [\psi_1(u) \pm \psi_2(u)] \ .
\eqno{(33)}
$$
Assume that the overall phase of 
both $\psi_0(u)$ and $\psi_1(u)$ has been chosen such that
they are both real-valued and both positive for $u\sim u_+$,
that is both $\psi_\iota(u)$ have positive amplitude in the
``right'' well of their double well potential $w(u)$. 
Clearly then $\psi_+(u)$ must be localized with most of its 
probability weight in the ``right'' well of $w(u)$, near $u=u_+$,
and likewise, by parity, $\psi_-(u)=\psi_+(-u)$ in the ``left''
well, near $u=u_-$.
At sufficiently high tunneling barrier, {\it i.~e.}
roughly when $\Omega_t\ll\Omega_h^\prime\equiv \eta_2-\eta_0$,
the spatial overlap between $\psi_+$ and $\psi_-$
in either well becomes negligibly 
small with $|\psi_+(u)|^2$ and $|\psi_-(u)|^2$
sharply peaked near $u_+$ and, respectively, near $u_-$.
We can thus estimate their $u$-matrix elements by a simple
tight-binding-like approximation, namely
$$
\langle \psi_\sigma | u | \psi_\sigma^\prime\rangle
\cong u_\sigma \delta_{\sigma,\sigma^\prime}
\eqno{(34)}
$$
for $\sigma,\sigma^\prime\in\{+,-\}$,
hence 
$$
\langle\psi_1|u|\psi_0\rangle\ =\ u_+\ =\ d_t/2 \ .
\eqno{(35)}
$$

Inserting the foregoing 
$
\langle\psi_1|u|\psi_0\rangle
$
into Eq.~(32) and the resulting $V_t$
into Eq.~(25), we can now calculate the tunneling contribution
to $\Sigma^{\prime\prime}$. This is greatly
simplified by the fact that $V_t$ is $\vec{q}$-independent,
{\it i.~e.}, the $\vec{k}^\prime$-integral can be
rewritten as a frequency integral over the interacting
electron density of states
$$
\rho(\omega)
\equiv\! {1\over N}\sum_{\vec{k}} 
{1\over \pi} (-)G^{\prime\prime}(\vec{k},\omega)\ .
\eqno{(36)}
$$
In the limit $T\!\to\!0$ and $|\omega|\lsim\Omega_t\ll\Omega_h^\prime$,
we get
$$
\Sigma_t^{\prime\prime}(\omega) \ \cong\ 
-{\pi\over 2}\ \Gamma_t\ \Theta\ (\Omega_t-|\omega|)
\eqno{(37)}
$$
where
$$
\Gamma_t = {1\over2} \rho(0) C^2 d_t^2\ .
\eqno{(38)}
$$

The gap-like depletion of spectral
weight in Eq.~(37), {\it i.~e.} the fact that
$\Sigma_t^{\prime\prime}(\omega)=0$ for electronic single-particle
energies $|\omega|$ up to $\Omega_t$ simply reflects the threshold
behavior associated with the density of states of the tunneling 
excitations: At $T=0$, an electron or hole quasi-particle
with an excitation energy $|\omega|$, measured from the
Fermi  level, cannot loose more than
$|\omega|$ in an inelastic emission or scattering 
process. Hence the electron or hole cannot undergo any such
inelastic processes if $|\omega|$ falls below $\Omega_t$,
which is the lowest bosonic excitation energy available for emission
into the lattice system. Also, at $T=0$, there are no 
thermal lattice excitations present to be absorbed by the quasi-particle.
Hence the damping $|\Sigma_t^{\prime\prime}|$
vanishes for $|\omega|<\Omega_t$.

Inserting Eq.~(37) into the Kramers-Kronig relation (27),
we find $\Sigma_t^\prime$ and, from that, the tunneling
contribution to the mass enhancement
$$ 
Z_t\ \equiv\  -\partial_\omega \Sigma_t^\prime (0)
\ \cong\ {\Gamma_t \over \Omega_t} \ .
\eqno{(39)}
$$
Since $\Gamma_t$ is comparable
to electronic energies, whereas $\Omega_t$ is small compared even
to typical phonon energies $\Omega_h^\prime$, we get $Z_t\!\gg\!1$.
In fact, in our model, Eq.~(6), we have very roughly
$$
\Omega_t\sim\Omega_h\exp(-c_t\Delta_B/\Omega_h)
\eqno{(40)}
$$
with $c_t$ of order unity.
From microscopic calculations of $W$,
we also know that $\Delta_B\!\sim\!E_P$
in the EP strong-coupling limit.{}{}{}{\RFCCa$^{,}$\RFCCb} 
Combining these results, we thus
recover quite naturally {\it{in our diagrammatic approach}}
the central result from strong-coupling-limit
polaron theory, namely{}{}{}{\RFCCa$^{,}$\RFCCb$^{,}$\RFHHa}
that the electron mass enhancement increases 
exponentially with $E_P/\Omega_h$, that is
$$
Z_t  
\sim  
{\Gamma_t \over \Omega_h} \exp(c_t E_P/\Omega_h) \ .
\eqno{(41)}
$$
The foregoing results, Eq.~(40),(41), apply
primarily in the strongly anharmonic limit
$\Delta_B\gg\Omega_h$. However, even in the intermediate
regime $\Delta_B\sim\Omega_h$, $\Omega_t$ and hence
$Z_t$ will still remain sensitively dependent on $\Delta_B$,
and, via $\Omega_h$ [Eqs~(9),(10)], on $M$ and $d_t$.

In the present diagrammatic model, 
the  $T\!=\!0$ mass enhancement is thus
caused directly by the
narrow, gap-like depletion in the self-energy 
spectrum discussed above.
The detailed spectral shape of $\Sigma^{\prime\prime}$ is of course
model-dependent. However, the gap-like
spectral weight depletion, over a very narrow
energy range $\sim\!\Omega_t$ around the Fermi level,
and the overall {\it{electronic}} magnitude of 
$|\Sigma^{\prime\prime}|\!\sim\!\Gamma_t\!\gg\!\Omega_t$ 
are generic features of the
lattice-mediated interaction $V_{AP}$ in the presence
of polaronic anharmonicity. Via the Kramers-Kronig relation, 
Eq.~(27), these features of $\Sigma^{\prime\prime}$ 
are directly translated into the
large $Z$ seen in $\Sigma^\prime$.

Another important feature of the tunneling mass 
enhancement $Z_t$ is its strong dependence 
on the isotope mass $M$ of the
lattice degrees of freedom. To see this, note first
that, all lattice-related model parameters 
such as $\Delta_B$, $d_t$ and $C$ and all bandstructure and spin
fluctuation-related model parameters, such as $t$, $g_s$, $\kappa_{AF}$,
$\Omega_s$ and $\mu$ must be regarded as being of purely electronic
origin. That is, in the general Born-Oppenheimer framework,
all these parameters do {\it not} depend sensitively on $M$. 
As a consequence, the damping strength $\Gamma_t$, from Eq.~(38),
is also $M$-independent.
However, $\Omega_{h}\propto M^{-1/2}$, 
as given by Eq.~(9), does depend on $M$. In contrast to this
rather modest power-law $M$ dependence of $\Omega_h$, the tunneling
splitting $\Omega_t$ is rather sensitive 
to changes in $M$. The ``exponential'' $M$-dependence 
of $\Omega_t$, reflected roughly in Eq.~(40), holds only
in the strongly anharmonic limit where $\Delta_B>\Omega_h$ and 
$\Omega_t\ll\Omega_h$. However, $\Omega_t$
will retain a substantial $M$-dependence even at weaker
anharmonicity where $\Delta_B\sim\Omega_h\gsim\Omega_t$. 
Hence, a strong isotopic mass dependence of $Z_t$ arises
in our model, via Eq.~(39), from the fact that the damping strength
$\Gamma_t$ is independent of $M$, whereas $\Omega_t$
is sensitively dependent on $M$. 

It is of interest to campare the foregoing results from the
anharmonic tunneling model to conventional harmonic phonon
exchange. To make this comparison meaningful, we specifically
ask what happens if we replace the anharmonic 
double-well $w(u)$, Eq.~(6), by a ``corresponding'' single-well
potential
$$
w_h(u) = {1\over2} K_h u^2
\eqno{(42)}
$$
whose the harmonic restoring force constant $K_h$ is chosen 
to coincide with the curvature of the double-well minima in $w(u)$,
{\it i.~e.}, $K_h=40\Delta_B/d_t^2$ as given by Eq.~(10) and the
corresponding harmonic phonon frequency 
$\Omega_{h}=\hbar(K_h/M)^{1/2}$ by Eq.~(9).
We also assume the same EP coupling constant $C$.

The derivation of the self-energy contribution $\Sigma_h$
and mass enhancement contribution from this harmonic Einstein
phonon exchange model is standard.{}{}{}{\RFGGa$^{,}$\RFGGb$^{,}$\RFJJ} 
The self-energy spectrum $\Sigma_h^{\prime\prime}$
has the same analytical form as $\Sigma_t^{\prime\prime}$
in Eq.~(37), with $\Omega_t$ replaced by $\Omega_h$ and 
$\Gamma_t$ replaced by a $\Gamma_h$, given analogous to Eq.~(38) by
$$
\Gamma_h =  {1\over 2} \rho(0) C^2 d_h^2
\equiv \rho(0) {C^2\over K_h} \Omega_h 
\eqno{(43)}
$$
Here the anharmonic tunneling fluctuation amplitude $d_t$ in Eq.~(38)
has been replaced by $d_h\equiv (2\Omega_h/\hbar K_h)^{1/2}$
which represents, in physical terms, the harmonic oscillator
zero-point fluctuation amplitude. An important point to note here
is that, again, $d_t$ in the double-well model is entirely a property
of the Born-Oppenheimer potential function $w(u)$, and hence purely
electronic in origin, 
{\it i.~e.} it does not depend on the isotopic mass $M$.
By contrast, $d_h$, and hence $\Gamma_h$ {\it do} depend on $M$.

Analogous to Eq.~(39), the harmonic phonon exchange
contribution to the self-energy is estimated by
$$
Z_h\equiv -\partial_\omega \Sigma_h^\prime (0)
\cong \Gamma_h/\Omega_h = \rho(0) {C^2\over K} \equiv \lambda_h
\eqno{(44)}
$$
which is of course just the conventional 
dimensionless Eliashberg EP coupling parameter $\lambda_h$,
written down here for the simple case of a harmonic Einstein model.
Comparing Eq.~(44) to Eq.~(41), we thus find that replacing
the harmonic Einstein oscillator well $w_h(u)$ by the double well
$w(u)$ increases the lattice contribution to the electron
mass enhancement by a factor
$$
{Z_t\over Z_h}\ =\ {d_t^2 \over d_h^2}
\  {\Omega_h\over\Omega_t}
\eqno{(45)}
$$
which will be large compared to unity, if the double well is wide,
$d_t\gg d_h$, and/or the tunneling excitation energy is small,
$\Omega_t\ll\Omega_h$.

An essential difference between the harmonic phonon
and the anharmonic tunneling exchange model
is their respective isotopic mass dependence which, in $Z_t$, is strong
whereas, in $Z_h$, is absent, since $\Omega_h$ cancels out in Eq.~(44).
The predicted isotope dependence of $Z_t$ can therefore 
be used as a  crucial experimental test
to establish or refute the participation of large-amplitude
lattice tunneling excitations in the low-energy electronic 
properties of the cuprates. 
Angle-resolved photoemission spectroscopy (ARPES)
measurements of the electron quasi-particle dispersion,
the electronic specific heat, the Drude weight
in the low-frequency optical conductivity and, in the
superconducting state, the low-temperature
($T\ll T_c$) London penetration depth
are possible candidates for exploring the isotopic mass dependence
of the electron mass enhancement in the cuprates.
Some experimental data for the isotopic mass dependence
of the penetration depth in one cuprate material already exist{}{}{}{\RFIsoExpZ}
and appear to indeed show a strong isotope dependence of $Z$. However,
we caution that these data{}{}{}{\RFIsoExpZ} are rather limited,
have not been reproduced independently at the present time,
and their interpretation is presently still controversial.{}{}{}{\RFIsoCritZ}

Indirect evidence for a strong isotope effect in $Z$
can be inferred from the extensively studied isotopic mass
dependence of the superconducting transition temperature $T_c$ in the 
cuprates,{}{}{}{\RFIsoExpTcA$^{,}$\RFIsoExpTcB} 
quantified in terms of the isotope exponent
$\alpha\equiv-\partial\log(T_c)/\partial\log(M)$ 
In several non-optimally doped cuprates systems,
values of $\alpha$ in the range $0.2-1.0$ are routinely observed,
even in samples having $T_c$-values as large as 
$40-60{\rm K}$.{}{}{}{\RFIsoExpTcA$^{,}$\RFIsoExpTcB} 

In theoretical studies of purely electronic AF spin fluctuation exchange
pairing models, giving rise to a $d_{x^2-y^2}$-pairing instability,
one finds{}{}{}{\RFPaScIsoD} that the inclusion of harmonic
phonon exchange suppresses $T_c$ primarily via the phononic contribution
$Z_h$ to the electron mass enhancement. 
As a consequence, the phononic $T_c$ suppression is largely
independent of the isotope mass, due to the absence of
isotopic mass dependence in $Z_h$, and the isotope exponent is small,
typically $|\alpha|<0.05$, in harmonic phonon
models. Large $|\alpha|$ values, $|\alpha|\sim 0.5-1.0$, 
could be produced by harmonic phonon exchange
only if the EP coupling in the model was made so strong that $T_c$ is 
suppressed to unrealistically low values ($< 10{\rm K}$).
By contrast, in spin fluctuation exchange $d$-wave pairing 
models with anharmonic tunneling exchange,
of the type described in the present paper,
large values of $\alpha$, of order $1.0$ and larger, 
could be easily reproduced {\it without} substantial
suppression of $T_c$.{}{}{}{\RFPaScIsoD}
Taken in conjunction with the experimental isotope
data,{}{}{}{\RFIsoExpTcA$^{,}$\RFIsoExpTcB} the model calculations suggests, that the
isotopic mass dependence of lattice contribution to 
$Z$ is indeed much stronger than coupling to 
a harmonic phonon system would allow to produce.

The foregoing conclusions{}{}{}{\RFPaScIsoD} apply to any
electronic ``excitation exchange'' pairing model,
with added coupling to harmonic phonon or, respectively, 
anharmonic lattice tunneling degrees of freedom,
and they go well beyond the specific framework 
of the AF spin fluctuation exchange mechanism. 
We caution however that the model isotope calculations{}{}{}{\RFPaScIsoD} 
{\it are} fundamentally based on the assumption 
that the pairing is primarily of electronic origin
and that it can be described within this general 
perturbative framework of an
electronic ``boson exchange'' theory. The latter
assumption is by no means firmly established
and or accepted at the present time.
More detailed experimental studies of the possible
isotopic mass dependence in the low-energy
electronic excitation spectrum of the cuprates
would therefore be exceedingly helpful in
establishing or refuting the participation
of anharmonic tunneling fluctuations in these
materials.

In interpreting such isotope experiments,
especially in the normal state of the cuprates,
it will be important to recognize that the
predicted strong isotopic mass dependence of $Z$
is limited to the low-temperature regime.
This is illustrated by Fig.~1 which shows
numerical results for the $T$-dependence of $Z$, 
obtained from Eq.~(31),
and also, separately, the anharmonic tunneling contribution
to $Z$, denoted by $Z_{AP}$, both calculated 
at the Fermi wavevector ${\vec{k}}_F$ along the $(1,1)$-direction,
where
$$
Z_{AP}(\vec{k}_F)\ \equiv\ -\partial_\omega \Sigma'_{AP}(\vec{k}_F,0)
\eqno{(46)}
$$
and $\Sigma_{AP}$ is the contribution to $\Sigma$
due to $V_{AP}$ in Eqs.~(22),(25). The calculation
was done with a set of model parameters which
is representative for the cuprates, namely
$\langle n_j\rangle=0.75$, that is a hole
doping concentration of $x\equiv1-\langle n_j\rangle=25\%$,
$t\!=0.35{\rm{eV}}$,{}{}{}{\RFBBa},
and spin fluctuation parameters{}{}{}{\RFFFa}
$g_s\!=\!1.06{\rm{eV}}$, 
$\Omega_s\!=\!0.105{\rm{eV}}$ and 
$1/\kappa_{AF}\!=\!3.33a$
with a lattice constant 
$a\!=\!3.8\AA$.
Also,{}{}{}{\RFCCa} 
$C\!=\!3.27{\rm{eV/\AA}}$,
$\Delta_B\!=\!75.29{\rm{meV}}$, 
$d_t\!=\!0.257{\rm{\AA}}$,
and 
$M\!=\!16{\rm{u}}$ 
for the lattice atomic mass, so that{}{}{}{\RFCCa$^{,}$\RFDD} 
$\Omega_t\!=\!10{\rm{meV}}$, 
$\Omega_h^\prime\!=\!50{\rm{meV}}$
and 
$\Omega_{h}\!=\!97.3{\rm{meV}}$.

For $T\ll\Omega_t$, we find numerically that 
$V_{AP}$, $\Sigma_{AP}$ and $Z_{AP}$ are indeed dominated by the
low-energy tunneling excitation $\iota=0\to\kappa=1$ [Eq.~(15)]
in $V_{AP}$. Hence, $V_{AP}$, $\Sigma_{AP}$ and $Z_{AP}$ are
essentially the same as, respectively, 
the $V_t$, $\Sigma_t$ and $Z_t$ discussed above. 
Our  analytical estimate $Z_t=\Gamma_t/\Omega_t$,
with $\Gamma_t$ from Eq.~(38) and the density of states
$\rho(\omega=0,T=0)\cong0.175t^{-1}$ from the inset of Fig.~2,
gives $Z_t\cong17.6$. This must be regarded as reasonable 
order-of-magnitude agreement with
the numerical $T=0$ result $Z_{AP}\cong 8.75$ in Fig.~1.
The analytical overestimate, by about a factor
of 2, is due to the fact that our estimate for the
tunneling matrix element 
$|\langle\psi_1|u|\psi_0\rangle|\cong d_t/2$,
entering into $\Gamma_t$, is an overestimate. 
Eq~(34) is strictly valid only
in the large-barrier limit $\Delta_B\gg\Omega_h$
which is not well realized here. Another factor affecting the
quantitative accuracy of our analytical estimates for
$\Sigma^{\prime\prime}_t$ and $Z_t$ is the assumption of an
approximately constant $\rho(\omega)$, on $\omega$-scales
of order $\Omega_t$. In the vicinity of the van Hove singularity
[see Fig.~2 inset and discussion below], this becomes a somewhat
crude approximation.

Along other directions of the Fermi wavevector, 
we find numerically quite similar
results. The overall dependence of $Z$ and $Z_{AP}$ on the
direction of $\vec{k}_F$ is weak. For example,
for ${\vec{k}}_F$ along the $(1,0)$-direction, $Z$ is about
$10\%-20\%$ larger than in the $(1,1)$ direction,
with quite similar $T$-dependence.

The important point to notice in Fig.~1
is that, while $Z_{AP}$ accounts for about $60\%$ of the total $Z$
at low $T$, $Z_{AP}$ is rapidly suppressed
when $T$ becomes comparable to the tunneling energy scale $\Omega_t$.
This happens because, upon raising $T$, the low-energy spectral gap in
$\Sigma_{AP}^{\prime\prime}$, Eq.~(37), gets "filled in," 
since thermally excited quasi-particles
can now emit and absorb thermal tunneling
excitations and thus cause damping, that is
$|\Sigma^{\prime\prime}|\!>\!0$ 
even at $\omega\!=\!0$.{}{}{}{\RFGGa$^{,}$\RFGGb$^{,}$\RFJJ} 
Already for $T\!\sim\!\Omega_t$, 
most of the $2\Omega_t$-gap in $\Sigma^{\prime\prime}$ 
and, along with it, most of the polaronic mass enhancement $Z_{AP}$
has  disappeared. 

Above this characteristic ``mass suppression'' temperature
scale, $T\!\gsim\!\Omega_t$, $V_{AP}$ contributes 
only a $T$- and $\omega$-independent damping to $\Sigma^{\prime\prime}$
which is physically quite similar to 
disorder scattering. The mass enhancement $Z$ is
then dominated by the spin fluctuation contribution
$V_{SF}$. In fact, as indicated by 
the dashed line in Fig.~1, for $T\gsim 2\Omega_t$, 
the mass enhancement in the full ``anharmonic phonon + spin fluctuation''
(APSF) model becomes indistinguishable from that of the
pure spin fluctuation (SF) model (without $V_{AP}$).
This $T$-independent damping contribution can also be estimated
analytically for $T\gg\Omega_t$. The result is roughly
$$
\Sigma_{AP}^{\prime\prime}(\omega)
\ \cong\ -{\pi\over 2} \Gamma_t\ \sim\ {\rm const}
\eqno{(47)}
$$
for all $\omega\ll t$.
Note that this is the same value
as  found for low $T$ at $\omega$ above the
excitation threshold, $\omega>\Omega_t$
in Eq.~(37). Note also that this damping contribution is
independent of the isotope mass $M$, as discussed above.

The foregoing high-$T$ results in the anharmonic
phonon model should also be contrasted 
with the conventional harmonic phonon exchange model.
In the latter, the mass enhancment contribution $Z_h$
becomes thermally suppressed when $T\gsim\Omega_h$.
However, the $T$- dependence of the electron quasi-particle
damping contribution $|\Sigma^{\prime\prime}_h|$ 
is quite different in the high-$T$ region, since{}{}{}{\RFGGa$^{,}$\RFGGb$^{,}$\RFJJ$^{,}$\RFAlleTr}
$|\Sigma^{\prime\prime}_h| \sim \lambda_h T$ rises linearly
with $T$. This difference in the $T$-dependence is generic to
the two types lattice dynamics models.
These different $T$-dependences arise
because in the harmonic model at $T>\Omega_h$, the
amplitude square of the fluctuating EP potential seen
by the electrons, $C^2\langle u_j^2\rangle$, increases
linearly with $T$, by the equipartition theorem.
By contrast, in the anharmonic tunneling model,
this amplitude square has a much weaker $T$-dependence,
since $C^2\langle u_j^2\rangle \sim C^2 d_t^2/4$ is 
determined, at both low and high $T$,
primarily by fluctuations between two wells,
not by fluctuations within a single well.

For comparion, we also quote the approximate,
Fermi surface averaged analytical results for the
spin fluctuation contributions to the self-energy,
$\Sigma_{SF}$, and to the mass enhancement, $Z_{SF}$.
Using standard Fermi surface averaging procedures,{}{}{}{\RFGGa$^{,}$\RFGGb$^{,}$\RFJJ}
we obtain from Eqs.~(17)-(19) and (25) 
for temperatures $T\ll\Omega_s$,
$$
\Sigma^{\prime\prime}_{SF}(\omega)
\ =\ - {\pi\over 4} \Gamma_s
{\rm min}\Big[{\omega^2\over\Omega_s^2},1\Big]
\eqno{(48)}
$$
and 
$$
Z_{SF}
\ =\ {\Gamma_s\over\Omega_s}
\eqno{(49)}
$$
where 
$$
\Gamma_s = {16\over\pi}
\rho(0) g_s^2 \langle F \rangle_{FS}
\eqno{(50)}
$$
and the Fermi surface average of
$F(\vec{k}-\vec{k}^\prime -\vec{Q}^*)$ from Eq.~(19),
denoted by $\langle F \rangle_{FS}$, 
is of order unity.
The low-energy spectral weight
depletion in $\Sigma^{\prime\prime}_{SF}(\omega)$
extends to $\Omega_s$. 
At excitation energies $\omega$ on the order
of the tunneling energy scale $\Omega_t$,
with $\Omega_t\ll\Omega_s$, 
the spin fluctuation contribution
to the damping is
$$
|\Sigma^{\prime\prime}_{SF}(\Omega_t)|
\ =\ {\pi\over4}\Gamma_s(\Omega_t/\Omega_s)^2
\ =\ {\pi\over4}{Z_{SF}\over Z_{AP} } {\Omega_t\over\Omega_s} \Gamma_t
\ll \Gamma_t\ .
\eqno{(51)}
$$
This is small compared to the lattice tunneling contribution
$|\Sigma^{\prime\prime}_{AP}(\Omega_t+0^+)|=(\pi/2)\Gamma_t$,
even though the respective mass enhancement contributions
$Z_{SF}$ and $Z_{AP}$ may be of comparable magnitude.

At very high temperatures, comparable to the spin fluctuation
energy scale $\Omega_s$, $Z_{SF}$ will be suppressed
by the thermal excitation mechanism 
in the spin fluctuation system,
analogous to the thermal suppression of $Z_{AP}$ described above.
Because of the much
higher energy scale $\Omega_s$, the  on-set of the thermal
suppression of $Z_{SF}$ is much more gradual, but it can already 
be clearly seen in the (compared to $\Omega_s$) relatively low temperature
regime $T\lsim 0.2t$ shown in Fig.~1. Given their large
$\Omega_s$ scale, this $Z_{SF}$ suppression 
effect may not be very important
in the cuprates, up to, say, $1000{\rm K}$.

\hbn
\centerline{\bf B. Coherent {\it vs.} incoherent polaron dynamics}

In spite of the large low-$T$ mass enhancement $Z_{AP}$,
the anharmonic lattice contribution to the 
damping in the high-$T$ regime is quite modest, of order
$|\Sigma_{AP}^{\prime\prime}|\cong (\pi/2)\Gamma_t\sim 0.28{\rm eV}$,
using the parameter set from Fig.~1 and Eq.~(38). 
Assuming a high-$T$ value $Z\sim 4$, say [see Fig.~1], 
this translates into an anharmonic tunneling contribution 
to the quasi-particle decay and transport relaxation rates of order
$$
{1\over \tau_{AP} } 
\ \cong \ 
{|\Sigma^{\prime\prime}_{AP}|\over Z} \ \sim\ 70{\rm meV}\ .
\eqno{(52)}
$$
This is roughly the right order of magnitude
magnitude, compared to typical
single-particle and Drude peak widths measured in the normal
state of the cuprates in the $200-400{\rm K}$ range.{}{}{}{\RFEE$^{-}$\RFEEb$^{,}$\RFDruOpt}
Recall here that the quasi-particle decay rates 
and transport relaxation rates, 
as reflected in the single-particle and 
Drude spectral (HWHM) peak widths, {\it are} being 
renormalized by the $1/Z$-factor, along with the quasi-particle
dispersion $E_{\vec{k}}$.

It is frequently argued that polarons cannot exhibit
coherent transport at temperatures exceeding some very
low ``de-coherence`` energy scale, set roughly by the
single polaron bandwidth, of order $8t/Z_P$ in our 2D model,
as estimated in the single-polaron strong-coupling theory.{}{}{}{\RFHHa$^{,}$\RFHHb}
Here, the strong-coupling polaron band width renormalization 
$Z_P$ can be easily as large as $10-10^4$ (or larger) 
and is exponentially dependent on the EP coupling strength 
and atomic mass, roughly as given in Eq.~(41).
According to this {\it single-polaron, strong-coupling} 
picture, polarons are inevitably 
condemned to incoherent propagation,
by hopping-type transport mechanisms,{}{}{}{\RFHHa} 
when $T$ becomes comparable to $8t/Z_P$.
Our foregoing diagrammatic results suggest otherwise:

The above estimated high-$T$ decay time, from Eq.~(52), 
and the resulting mean free path, are still
about an order of magnitude away 
from the Mott-Ioffe-Regel limit, {\it i.~e.} well
on the coherent side of that limit.
The quasi-particles, albeit damped,
remain coherent at high-$T$, due to the fact that
the polaronic mass enhancement $Z_{AP}$ is being suppressed
at $T\gsim\Omega_t$ and due to the fact that
the high-temperature ``remnants'' of the low-$T$ anharmonic 
tunneling fluctuations do not scatter the quasi-particles
very strongly. At least for the specific parameter set
used above,  the scattering from the 
``remnant'' anharmonic fluctuations
is by no means sufficient to drive the electron quasi-particles
into the incoherent / hopping type regime.
The physical picture that emerges here is that,
at $T\gsim\Omega_t$, only the coherence of the 
lattice tunneling groundstate -- but {\it not} the coherence of
the electronic quasi-particles -- is being lost when
$T$ reaches the {\it lattice} tunneling energy scale.

How can these two seemingly quite contradictory physical pictures
be reconciled ? The answer, we believe, lies in the 
parameters. The above-described strong-coupling small
polaron picture is based on the EP strong-coupling
limit where, in the underlying microscopic EP model,
the EP coupling strength $E_P\equiv C^2/K$
exceeds the bare bandwidth $8t$ and, 
at the same time, the tunneling barrier
heights  $\Delta_B$ of the polaronic 
Born-Oppenheimer {\it lattice} potential
$W$ and the polaron binding energies
(relative to the delocalized carrier state)
are large, namely, of order $E_P$.

In this limit the polaronic lattice fluctuation amplitude
$d_t$ is of the order{}{}{}{\RFCCa$^{,}$\RFCCb} $d_t\sim C/K$, hence $Cd_t\sim E_P$,
and in our local double-well lattice model, Eq.~(6), 
the condition $E_P\gg t$, via Eq~(38), translates into
$$
\Gamma_t \sim {1\over 2}\rho(0) E_P^2 \sim {E_P\over 8t} E_P \gg t\ .
\eqno{(53)}
$$
We are assuming here 
$\rho(0)\sim 1/8t$ so that $\rho(0)E_P\gsim {\cal O}(1)$.
Under these conditions, our diagrammatic approach
breaks down at $T\gsim\Omega_t$,
but, nevertheless, the diagrammatics 
signals its own inadequacy,
as the quasi-particle damping, estimated from Eq.~(47), 
will indeed approach or 
exceed the Mott-Ioffe-Regel limit.

This strong-coupling picture is {\it inevitably} correct
for low-density polaron systems with a nearly
empty conduction band and with short-range EP coupling
in spatial dimensions $D\ge 2$.
In this type of EP system, the conditions
$\Delta_B\sim E_P\gg t$ must be satisfied in the polaronic
regime because, unless $E_P>E_P^{(crit)}\gg t$, 
polarons will simply not be formed.{}{}{}{\RFHHxb}
Here, $E_P^{(crit)}$ denotes the minimum EP coupling strength
needed to stabilize the polaronic self-localized 
carrier state against delocalization. In the nearly empty-band
situation, $E_P^{(crit)}$ is of order of the bandwidth or larger.{}{}{}{\RFCCa$^{,}$\RFHHxb}

However, in the case of the Holstein-Hubbard (and related) models,
as applied to the cuprates, polaron formation arises in
the near $1/2$-filled band limit and the polaronic
carriers are dopant induced holes 
in  a strongly correlated Mott-Hubbard insulator electron background.
Extensive microscopic calculations{}{}{}{\RFCCa$^{-}$\RFCCd} have shown that the
presence of AF spin correlations in the near $1/2$-filled
Hubbard electron system drastically reduces the threshold for
polaron formation in such a system, roughly to $E_P^{(crit)}\sim2-3t$
in realistic parameter regimes.
Furthermore, for $E_P$ near $E_P^{(crit)}$, the lattice potential
barriers separating different polaronic minima can become
quite small, so that $\Delta_B\ll t$ or even $\Delta_B\sim\Omega_h$
and also $Cd_t< E_P$. In this limit, we get
$$
\Gamma_t\ <\ {1\over 2} \rho(0) E_P^2\ \lsim\ t
\eqno{(54)}
$$ 
In this `` weakly bound polaron'' regime, 
a diagrammatic treatment based on
a coherent quasi-particle picture is likely to become applicable, 
even for  $T\gsim\Omega_t$. On the other hand, the strong-coupling 
expansion, based on the assumption $E_P\gg t$, is likely to fail here.

It is conceivable that microscopic EP models may exhibit a cross-over
from the strongly bound regime, Eq.~(53), to the weakly bound
regime as a function of increasing doping and/or increasing
temperature. Such a cross-over can be driven by mutual
screening of the attractive polaronic EP potential wells. This
screening occurs at finite polaron density when nearby polaronic 
self-localized wavefunctions and EP potential wells
begin to overlap. It has the effect of lowering the height $\Delta_B$
and the width $d_t$ of the polaronic inter-site tunneling barriers 
in the lattice potential $W$. 
This barrier screening effect has recently been demonstrated
by microscopic calculations for the Holstein-Hubbard
model.{}{}{}{\RFCCb} Thermal
de-trapping of the polaronic carrier from its EP potential
well will also tend to enhance this barrier screening
effect by incrasing the population of 
unbound carriers. Hence, raising the temperature
can have qualitatively the same effect as raising the doping level,
namely, to push the system towards weaker
polaronic binding energies and lower barrier heights. 
It will be interesting to investigate
whether such a cross-over is realized in at least some of 
the cuprates which, as a function of doping and temperature
seem to show evidence of hopping-type conductivity
at low doping and low $T$, but exhibit apparently coherent
transport at larger doping and/or higher $T$.

The partially ionic character of the cuprates and resulting
spatially extended Fr\"ohlich-type EP couplings are another
factor which may contribute to the existence of a ``weakly bound''
polaron regime. Such long-range EP couplings have the effect
of lowering the overall threshold $E_P^{(crit)}$ by allowing
``large'' ({\it i.~e.} spatially extended) 
polaron formation to occur even in dimension $D\ge 2$.{}{}{}{\RFHHxb$^{,}$\RFEmBP}
Extensive phenomenological studies have so far
considered only large bi-polarons and only in the context of continuum
models.{}{}{}{\RFHHxb$^{,}$\RFEmBP} It will be of considerable interest to investigate
such longer-range EP coupling with discrete crystal lattice
models and to explore the cross-over from large / weakly bound
to strong-coupling / strongly bound polaron behavior.

\hbn
%
\centerline{\bf C. Density of states and single-electron spectra}

In Fig.~2, we show the resulting density of states,
$\rho(\omega)$, calculated for the same parameter set
as in Fig.~1, from Eq.~(36), at $T=0$ and at 
$T=0.029t\cong\Omega_t$.
The most prominent feature in Fig.~2 is the narrow,
strongly $T$-dependent peak in close proximity
to the Fermi level $\omega\!\equiv\!0$ in
$\rho(\omega)$, as
shown by the full line in Fig.~2.{}{}{}{\RFLL}
At $T\!=\!0$, we can show
that $\rho(\omega)$ near the Fermi level is essentially just a
narrowed image of the non-interacting density of states{}{}{}{\RFJJ}
$\rho^{(0)}(\omega)$ for the same band filling, that is, for 
$|\omega|\!\lsim\!\Omega_t$
$$
\rho(\omega) 
\cong \rho^{(0)}(\bar{Z}\omega)
\eqno{(55)}
$$
with $\bar{Z}$ denoting the ${\vec{k}}_F$-average of $Z$. 
The ``resonance'' peak in Fig.~2 is thus primarily
a narrowed image of the non-interacting 2D van Hove singularity
and its peak position $\omega_{vHs}$ relative to the Fermi
level $\omega=0$ is roughly given by
$$
\omega_{vHs} = {\omega_{vHs}^{(0)}\over\bar{Z}}
\eqno{(56)}
$$
where $\omega_{vHs}^{(0)}$ denotes the position of 
the van Hove singularity, relative to the Fermi level,
in the corresponding non-interacting system at the same
band filling.

A rough derivation of this result, Eq.~(55), can be sketched 
as follows: First, neglect the $\vec{k}$ dependence of
$\Sigma$ around the Fermi surface and replace 
$\Sigma(\vec{k},\omega)$ by an appropriate Fermi surface
average $\bar{\Sigma}(\omega)$.{}{}{}{\RFGGa$^{,}$\RFGGb$^{,}$\RFJJ} 
Second, neglect the small
imaginary part $\bar{\Sigma}^{\prime\prime}$ and 
approximate its real part by the standard 
``Fermi liquid'' approximation
$$
\bar{\Sigma}^\prime(\omega)\cong
(1-\bar{Z})\omega + {\cal O}(\omega^2)
\eqno{(57)}
$$
where $\bar{Z}\equiv 1-\partial_\omega\bar{\Sigma}(0)$.
Third, insert this into Eqs.~(26) and the resulting $G^{\prime\prime}$
into Eq.~(36) to arrive at Eq.~(55).

Note, in Fig.~2, that the logarithmic divergence 
at the van Hove singularity
has been slightly smeared out, even at $T=0$, by the
damping ($|\Sigma^{\prime\prime}|\!>\!0$)
at finite $\omega$ and by the $\vec{k}$ dependence
of $\Sigma$ along different directions on the Fermi surface.
These damping and $\vec{k}$-dependence effects are neglected
in the derivation of Eq.~(55).
Also, as shown in the inset of Fig.~2, 
the resonance peak is rapidly suppressed, along with $Z(T)$,
when $T$ becomes comparable to $\Omega_t$.
This is again caused by the strong thermal
damping at $\omega\!=\!0$ 
and by the thermal suppression of $Z(T)$, 
when $T$ bcomes of order $\Omega_t$
or larger. Because of the thermal damping effects,
the density of states peak in the ``APSF'' model (including $V_{AP}$)
is suppressed faster with increasing temperature than in the pure
``SF'' model (without $V_{AP}$, see dashed line in Fig.~2 and inset); 
the APSF model generates an additional strong
damping already when $T\sim\Omega_t$ which is absent in the SF model.

A careful study of the doping dependence of 
the density of states peak near $\omega=0$
reveals that, while substantially contributing to this peak, 
the van Hove
singularity in the non-interacting density of states $\rho^{(0)}$ 
is not necessarily required to generate such a $\rho(\omega)$-peak.
Even for a completely flat $\rho^{(0)}(\omega)$, {\it e.~g.}
at band fillings where the van Hove singularity is far
away from Fermi level $\omega=0$, it is possible to obtain
a pronounced peak structure in $\rho(\omega)$ near the Fermi level.
The general conditions under which this happens are
that the self-energy spectrum $|\Sigma^{\prime\prime}(\vec{k}_F,\omega)|$ 
on the Ferrmi surface
$\vec{k}_F$ must (i) be very small (or zero) at $\omega=0$
and (ii) rise very rapidly to large values on some
very low frequncy scale, denoted by $\Omega$, such that 
$$
|\Sigma^{\prime\prime}(\vec{k}_F,\Omega)|\sim t\gg
\Omega,|\Sigma^{\prime\prime}(\vec{k}_F,0)|\ .
\eqno{(58)}
$$
Note that these conditions can be satisfied in our
APSF model at temperatures $T\ll\Omega_t$, 
where $\Omega_t$ represents the low-energy
scale $\Omega$ and it is therefore possible to get
$\Sigma^{\prime\prime}_t(\Omega_t+0^+)\sim\Gamma_t\gg\Omega_t$.
By contrast, these conditions can, realistically, {\it not}
be satisfied in harmonic phonon exchange models, as exemplified by
the harmonic Einstein model discussed above.
In the latter, we have, from Eqs.~(43) and (44), 
$\Sigma^{\prime\prime}_h(\Omega_h+0^+)\cong (\pi/2)\lambda_h\Omega_h$,
which is comparable to the phonon energy scale
for typical Eliashberg $\lambda_h$-values of order unity.
Hence, EP-induced density of states peak effects
are another important distinguishing feature
between harmonic phonon and anharmonic tunneling exchange.
We will now describe how this density of states peak effect
arises.{}{}{}{\RFMM}

Under the conditions of Eq.~(58), the quasi-particle peak in 
$G^{\prime\prime}$ is rapidly damped out with increasing
quasi-particle energy which, in essence, has the effect 
of pushing spectral weight away from the Fermi level
for finite $\omega$, of order $\Omega$. This spectral weight
removal, due to strong damping at $\omega\sim\Omega$,
occurs {\it in addition} to the spectral weight removal
caused by the mass enhancement factor $Z$.
In contrast to the latter, the former spectral weight
removal is not being compensated
for by the reduced quasi-particle dispersion,
when one averages $G^{\prime\prime}$ over all $\vec{k}$
to calculate $\rho(\omega)$.
At $|\omega|\ll\Omega$, on the other hand, spectral
weight in $G^{\prime\prime}$ is removed only by the $Z$-effect
which, in $\rho(\omega)$, {\it is} being compensated
for by the reduced quasi-particle dispersion.
As a consequence, 
spectral weight 
will be removed from $\rho(\omega)$ for finite $\omega$,
of order $\Omega$, but not at very low $\omega$, well below $\Omega$.
It is important to realize that, therefore, the resulting
peak in $\rho(\omega)$ at $\omega=0$ does {\it not} 
arise from an enhancement of $\rho(\omega)$ {\it at} the Fermi level,
but rather from a suppression of $\rho(\omega)$ {\it off} the Fermi
level. In fact, neglecting the $\vec{k}$-dependence of $\Sigma$,
the density of states right at the Fermi level is essentially unchanged
by the interaction, that is, from Eq.~(57)
$$
\rho(0)\cong\rho^{(0)}(0)\ .
\eqno{(59)}
$$

Notice also that a $\rho(\omega)$-peak, generated solely by the foregoing
damping-induced spectral weight removal mechanism from a flat non-interacting
density of states ({\it i.~e.} from a $\rho^{(0)}(\omega)\cong{\rm const}$
at $|\omega|\sim \Omega$), will be centered right
at the Fermi level $\omega=0$. As shown in the inset in Fig.~2,
this is not the case in our $T=0$ model calculation. Here, the narrowed
image of the van Hove singularity, centered at an $\omega_{vHs}\cong0.02t$,
is below $\Omega_t$ and 
superimposed on the damping-induced spectral weight suppression
which occurs mainly for $|\omega|>\Omega_t$ in our model, due to
the almost completely gapped $\Sigma^{\prime\prime}$ 
at $|\omega|<\Omega_t$, Eq.~(37). 
However, if we change the band-filling so as to move
the van Hove singularity further away from the Fermi level,
then the narrowed van Hove image moves out to
larger $\omega_{vHs}$ and gradually disappears when $\omega_{vHs}$
becomes comparable to $\Omega_t$ while, at the same time,
the damping-induced peak at $\omega=0$ emerges, abeit generally
with less pronounced peak height.

We emphasize the foregoing especially in relation to
photoemission studies of the $\vec{k}$-integrated
density of states and its doping evolution
in the cuprates. The foregoing considerations suggest that
one needs to be very careful in the interpretation of observed
$\vec{k}$-integrated density of states peaks. Such peaks
need not necessarily be a manifestation
of van Hove singularities, but may rather be caused by
other effects, such as strongly frequency dependent damping
near the Fermi level.

Also shown in Fig.~2, by the dot-dashed line,
is the result of a non-self-consistent 1st order SF model calculation
for $\rho(\omega)$. Here, the fully dressed internal single-electron
Green's function in Eqs.~(22), (25) is replaced
by the bare Green's function at the same band filling, that is,
written in the Matsubara domain, with
$$
G^{(0)}(\vec{k},i\nu)=
{1\over i\nu-\epsilon_{\vec{k}}^{(0)} }\ .
\eqno{(60)}
$$
Here $\epsilon_{\vec{k}}^{(0)}$
denotes the non-interacting band, obtained from Eq.~(24)
with the non-interacting
value of the chemical potential $\mu^{(0)}$ at the same band filling
$\langle n_j\rangle^{(0)}=0.75$.

Similar 1st order calculations were first reported
for this SF model in Ref.~[\olRFFFa]. An important prediction
from these early 1st order model calculations was the demonstration
that the spin fluctuation exchange causes a pronounced ``pseudo-gap''
to open up in the density of states near the Fermi level.
This pseudo-gap, {\it i.~e.} a strong 
suppression in the density of states near $\omega=0$, is 
indeed also present in our 1st order calculation, with the
spectral weight depletion extending approximately
from $\omega\sim0$ up to $\omega\sim 3.8t$ in Fig.~2.
However, in our self-consistent SF calculation
(using a fully dressed $G$ as the internal electron line),
the pseudo gap has largely disappeared, as shown by the dashed line
in Fig.~2. In the $\omega$-region of the gap, where the 1st order
$\rho(\omega)$ is being depleted, the self-consistent
$\rho(\omega)$ is almost flat. 
Hence, it is not quite clear whether the original
claim of pseudo-gap generation in AF
spin fluctuation exchange exchange models
is really tenable. At the very least, this 
spin fluctuation induced pseudo-gap effect, 
if present at finite doping, appears to be quite
model dependent, that is,
very sensitive to the level of approximation employed
in the calculation and, possibly, to the choice
of model parameters. 

In Fig.~3, we show the single-hole excitation spectra{}{}{}{\RFJJ}
$$
P({\vec{k}},\omega)\!\equiv\!\pi^{-1}
|G^{\prime\prime} ({\vec{k}},\omega)|f(\omega),
\eqno{(61)}
$$
with the Fermi factor $f(\omega)\equiv 1/(e^{\beta\omega}+1)$,
for $T\!=\!0.029t\!\cong\!\Omega_t$
and several ${\vec{k}}$-points along the $(1,1)$ direction.
At all ${\vec{k}}$, $P({\vec{k}},\omega)$ exhibts a strong
continuous "incoherent" background which is almost constant, 
down to $\omega\!\sim\!-2\;t$, and extends up to the Fermi level 
$\omega\!\equiv\!0$ for ${\vec{k}}\!=\!{\vec{k}}_F$.
As ${\vec{k}}\!\to\!{\vec{k}}_F(1,1)$,
the expected quasi-particle peaks at energies $\!E_{\vec{k}}\!<\!0$
emerge from this background, 
sharpen, and disperse towards $E_{\vec{k}}\!=\!0$.
These results are quite similar to typical 
angle-resolved photoemission spectroscopy
(ARPES) data obtained in the cuprates at $T\!>\!T_c$.{}{}{}{\RFEE$^{-}$\RFEEb$^{,}$\RFNN}
Notice that, at the temperature $T=0.029t$ used here
({\it i.~e.} $T\sim 110{\rm K}$, assuming $t\!=0.35{\rm{eV}}$),
the lattice contribution to the mass enhancment $Z_{AP}$ is already
noticeably reduced by the thermal $Z$-suppression.
Hence, except for the additonal line broadening caused
by $\Sigma^{\prime\prime}_{AP}$, the shapes and the dispersion
of the quasi-particle peaks are already very similar
to those in the purely electronic SF model, without
coupling to lattice degrees of freedom.
We emphasize that the normal-state ARPES line shapes
and quasi-particle dispersion
in the cuprates are measured at typical $T\!>\!T_c\!\sim\!80-120\;^o\!K$.
Even though coupling to lattice tunneling excitations
may be important in shaping the quasi-particle dynamics
of the cuprates at low temperatures, 
the tunneling excitations do not necessarily have
any pronounced signature at normal state temperatures
which may well be in the regime $T\gsim\Omega_t$.

The quasi-particle energy band $E_{\vec{k}}$, 
shown in Fig.~4, also resembles typical cuprate
ARPES data near ${\vec{k}}_F$.{}{}{}{\RFEEb} In particular,
the very flat ("heavy mass") portions in $E_{\vec{k}}$ near 
${\vec{k}}_F$ along the $(1,0)$-direction
are very similar to recently reported quasi-particle 
dispersion curves along certain ${\vec{k}}$-directions 
in the cuprates.{}{}{}{\RFEEb}
In our model, these very "heavy" sections in $E_{\vec{k}}$
are due to the close proximity of ${\vec{k}}_F(1,0)$
to the non-interacting band saddle point, the van Hove point $(\pi,0)$,
with substantial additional band "flattening"
caused by the mass enhancement $Z$. 
Note here that Eq.~(29) 
has multiple solutions for $E_{\vec{k}}$ 
near ${\vec{k}}_F$, as shown in Fig.~4.
The $E_{\vec{k}}$-branch crossing the Fermi level
$E_{{\vec{k}}_F}\!\equiv\!0$ corresponds to the 
quasi-particle peak in $G^{\prime\prime}({\vec{k}},\omega)$.
Other branches do {\it{not}} necessarily produce well-defined 
separate peaks in $G^{\prime\prime}$, because of damping.
\hbn
\hbn
\hbn
\centerline{\bf IV. CONCLUSION}

In summary, we have proposed a highly simplified
anharmonic lattice potential model to study
the effect of polaronic tunneling excitation
and anti-ferromagnetic (AF) spin fluctuation exchange
on the quasi-particle dynamics in a 2D
single band model. The effects of both polaronic lattice tunneling
and AF spin fluctuation exchange are treated in a self-consistent
Migdal-type single-exchange approximation for the single-particle
{}self-energy.
At temperatures $T$ and excitation 
energies $\omega$ which are well below the anharmonic 
tunneling excitation energy scale $\Omega_t$,
the single-electron self-energy is dominated by the
exchange of lattice tunneling excitations 
which give rise to a large and strongly isotopic mass dependent
electron mass renormalization $Z$.

With increasing $T$, the lattice tunneling 
contribution to the electron mass enhancement,
along with the isotope effect of $Z$, are rapidly suppressed
and they both essentially disappear when $T$ exceeds $\Omega_t$.
In the high temperature regime $T\gsim 2\Omega_t$,
the electron mass enhancement $Z$ is completely
dominated by the AF spin fluctuation exchange;
the lattice tunneling excitations contribute only to the
electron quasi-particle damping $|\Sigma^{\prime\prime}|$
and this damping contribution is largely $T$-independent 
and independent of the isotope mass. 
This high-temperature damping effect is therefore
quite similar -- and experimentally indistinguishable from 
-- conventional disorder scattering.

Surprisingly, this anharmonic lattice contribution to the
quasi-particle damping at high temperatures $T\gsim\Omega_t$
can be quite small compared to the electronic bandwidth, even 
if the anharmonic lattice tunneling fluctuations 
produce a very large mass enhancement at low $T$. 
The relatively small damping at high $T$
allows the electronic quasi-particle propagation to remain
coherent at high temperatures, despite of the 
polaronic character of the underlying lattice dynamics.
We have described in detail the ``weakly bound polaron''
conditions under which this behavior may occur, within the
framework of a microscopic Holstein-Hubbard-type model;
and we have contrasted it with the fundmentally incoherent
high-$T$ behavior in ``strongly bound'' polaron systems. 

In comparing the isotopic mass dependences of the electron
mass enhancement $Z$ at low $T$
in the anharmonic tunneling exchange model 
{\it versus} a conventional phonon exchange model,
we find that the two types of models exhibit
fundamentally different behavior:
The $Z$ contribution in the anharmonic tunneling 
exchange model {\it is} isotope dependent,
whereas, in the harmonic phonon model, it {\it is not}.
Furthermore, in contrast to the $T$-independent 
damping in the anharmonic tunneling exchange model
at high $T$, the high-$T$ damping contribution 
in the harmonic phonon model increases linearly with $T$,
In this context the demarkation between ``low $T$'' and ``high $T$''
is given, respectively, by the tunneling energy scale $\Omega_t$
in the former and by the harmonic (Einstein or Debye)
phonon energy scale in the latter model.

Based on the foregoing results, we suggest to 
explore the isotopic mass dependence of the low-$T$
electron mass enhancement experimentally,
in order to establish 
whether or not anharmonic lattice tunneling excitations
contribute in a significant manner to the low-energy
electronic properties in the cuprate materials.
Possible candidate experiments for pursuing such 
investigations are
quasi-particle energy dispersion curves, as measured
in angle-resolved photoemission spectroscopy (ARPES);
Drude spectral weights, as observed in the optical conductivity;
electronic specific heat; and, in the superconducting phase,
the $T\to0$ London penetration depth.

In the $\vec{k}$-integrated density of states,
low-$T$ tunneling excitation exchange
can produce a narrowed image of the near-Fermi-level
features of the non-interacting density of states, such as, for example, 
a narrowed van Hove singularity peak. 
At finite temperatures $T\sim\Omega_t$, the van Hove peak
is completely suppressed due to the thermal damping.
Likewise, the quasi-particle dispersion $E_{\vec{k}}$
is largely unaffected by the anharmonic tunneling excitation
exchange and the single paricle spectral function, as observed
in ARPES experiments, is affected only by the $T$-independent
broadening, when $T\gsim\Omega_t$.

We note that our results concerning the low-$T$
narrowed imaging of the van Hove peak and 
its thermal suppression 
are generic to {\it{any}} strong-coupling low-energy boson
exchange model, as demonstrated by our results
for the purely electronic spin fluctuation 
exchange model (see Fig.~2).
In boson exchange models of the type discussed here,{}{}{}{\RFOO}
the rapid thermal suppression of the
narrowed van Hove peak in $\rho(\omega)$
is ubiquitous and entirely in line with the finding 
that van Hove singularities alone are not sufficient
to bring about or substantially enhance 
high-$T_c$ pairing instabilities.
Our results thus strongly support the notion{}{}{}{\RFOO} 
that a conjectured "van Hove scenario"{}{}{}{\RFPP}
of high-$T_c$ superconductivity does not exist.

To the extent that the cuprates can be described 
in terms of this type of low-energy ``tunneling boson'' 
exchange picture,
we emphasize that the observed normal-state mass enhancements{}{}{}{\RFEE$^{-}$\RFEEb}
$Z(T)\!\sim\!2-4$ may substantially underestimate
the true bosonic coupling strength{}{}{}{\RFJJ},
$\lambda\!\cong\!Z(T\!=\!0)-1$. As shown here,
$Z(T)\!\ll\!Z(T\!=\!0)$
if the normal state measurement temperature $T$ 
({\it e.~g.} $T\!>\!T_c\!\gsim\!80-120K$ 
in the cuprates !) exceeds 
the relevant boson energy scale $\Omega$.
Also, for $T\!\ll\!\Omega$, the full $Z(T\!=\!0)$ will be observable
only within a very narrow energy range ($\ll\!\Omega$) 
around the Fermi level, which could well be below current
limits of resolution.{}{}{}{\RFEE$^{-}$\RFEEb}
In high-$T_c$ systems where $\Omega_t\lsim T_c$
for $T\!\ll\!T_c$, it needs to be explored to what
extent the full normal-state value of 
$Z(T\!=\!0)$ is observable
in the presence of a superconducting gap $\Delta$ 
with $\Delta\!\gsim\!\Omega$.
Finally, we note that a strongly $T$-dependent $Z$ has important 
implications for the $T$-dependence of quasi-particle 
lifetimes $\tau$, since $1/\tau\!\cong\!|\Sigma^{\prime\prime}|/Z$ {\it{is}} 
renormalized by $Z$.

Given sufficient energy resolution,
the full $T\!=\!0$ normal-state mass enhancement 
{\it{should}} be observable in systems where $T_c\!\ll\!\Omega$, 
at temperatures $T$ well below $\Omega$.
From the $T$-dependence of $Z(T)$, the relevant
boson energy scale can then be established.
Also, the possible observation of 
the $\rho(\omega)$-resonance peak should be explored.
The resonance is observable 
by either direct or inverse angle-integrated
photoemission spectroscopy, depending on the resonance
peak position $\omega_{vHs}$ relative to the Fermi level $\omega\!=\!0$.
The resonance can also cause
a low-energy gap structure, on an energy scale of order 
$|\omega_{vHs}|$, in the 
two-particle electronic response functions.

To put our model results in a more general
perspective, we note that
the specific shape of the spectral weight depletion
around the Fermi level $\omega=0$ in the self-energy
[see Eq.~(37)]
is obviously model dependent. However, the
qualitative feature of its existence
is characteristic of the low-energy behavior of 
any Fermi liquid. What is unusual about this
Fermi liquid behavior in the lattice tunneling self-energy
studied here, is the fact that the Fermi liquid
character begins to break down already at a 
very a low excitation energy ($\omega$) and
temperature scale, set by the tunneling energy $\Omega_t$.
This can, but need not be lower than all other 
electronic or lattice energy scales, because of the great
sensitivity of tunneling matrix elements to small changes
in the tunneling barriers. Thus, the tunneling energy scale 
(i) could vary substantially between different materials;
(ii) within the same parent material, it could vary substantially
as a function of doping level; and 
(iii) it could well be lower than either the harmonic phonon
or even the superconducting $T_c$ scale in the cuprates.

These last features of our model are 
generic to the low-energy physics of 
strongly correlated polaronic electron
systems; they do not depend on the model 
simplifications
we have introduced here. What {\it is} quite likely to change
in a more realistic polaron model is the
structure of the anharmonic lattice dynamics, 
as reflected {\it e.~g.} in the $\omega$- and 
$\vec{q}$-dependence of $V_{AP}$. This, in turn,
will have important ramifications for
the details of the $T$- and $\omega$-dependence of the self-energy
in such systems.

Lastly, we should also emphasize that the Hubbard-type 
local Coulomb correlations play a central role 
in such a polaronic picture, for two reasons.
Firstly, the local Coulomb repulsion is important
to prevent the formation of bi-polarons and hence,
to preserve the essentially fermionic character
of the quasi-particles.
Secondly, the Coulomb correlations are the
origin of the AF spin correlations in the nearly
$1\over2$-filled regime. As discussed previously,{}{}{}{\RFCCa$^{-}$\RFCCd} 
the AF spin correlations are essential to allow 
doping induced carriers to form polarons at
much weaker EP coupling strengths than would be required
in weakly correlated systems. As a consequence,
the resulting polaronic system can have much lower tunneling
barriers and hence, reach much {\it larger} $\Omega_t$-scales than
one would estimate for weakly / un-correlated small-polaron systems.

Note that the latter is crucial if one wants to apply
such a polaronic picture to the cuprates. On the one
hand, one will {\it need} low tunneling energy scales in order to
explain the breakdown of Fermi liquid behavior down to very
low temperatures, down to, say, $\sim 10$K, in some cuprate systems; 
on the other hand, one will need large tunneling energy
scales to to explain the unusual $T$-dependences
of the quasi-particle and transport dynamics which are found
over very large temperature scales, up to $\sim 100-1000$K.
A lattice excitation spectrum covering 
several orders of magnitude of excitation energy will therefore
be essential. This, again, emphasizes the importance
of more realistic models of the polaronic lattice
dynamics and will require further study.
\hbn
\hbn
\hbn
\centerline{\bf ACKNOWLEDGMENT}

We would like to acknowledge discussions with
J.W. Allen, A. Arko, M. Jarrell, A. Kampf, 
M. Norman, C.G. Olson, R.J. Radtke,
D.J. Scalapino, J.R. Schrieffer and Z.-X. Shen.
This work was supported by the National Science Foundation under Grant No. 
DMR-9215123, by the University of Georgia
Office of the Vice-President for Research, and by the University
of Georgia computing facilities.
\hfil
\noindent
\vfil
\eject
\centerline{REFERENCES}
\hbn
\item{[\olRFAA]} 
  P.W. Anderson,
        Science {\bf{235}}, 1196 (1987);
  F.-C. Zhang and T.M. Rice,
        Phys. Rev. B {\bf{37}}, 3759 (1988).
%
\item{[\olRFBBa]} 
      H.-B. Sch\"uttler and A.J. Fedro,
        Phys. Rev. B {\bf{45}}, 7588 (1992).
\item{[\olRFBBb]} 
      M.S. Hybertsen et al., 
        Phys. Rev. B {\bf{41}}, 11068 (1990).
\item{[\olRFBBc]} 
      A.K. McMahan et al., 
        Phys. Rev. B {\bf{42}}, 6268 (1990).
\item{[\olRFBBd]} 
      S.B. Bacci et al.,
        Phys. Rev. B {\bf{44}}, 7504 (1991).
%
\item{[\olRFCCa]} 
      J. Zhong and H.-B. Sch\"uttler,
        Phys. Rev. Lett. {\bf{69}}, 1600 (1992).
\item{[\olRFCCb]} 
      K. Yonemitsu, J. Zhong and H.-B. Sch\"uttler, 
        preprint cond-mat/9805320;
      H.-B. Sch\"uttler {\it et al.},
      J. Supercond. {\bf 8}, 555-558 (1995).
\item{[\olRFCCc]} 
K. Yonemitsu et al., 
Phys. Rev. Lett. {\bf 69} 965 (1992).     
\item{[\olRFCCd]}
H. R\"oder {\it et al.}, \prb {\bf 47}, 12420 (1993); 
H. Fehske {\it et al.},
J. Phys. Condens. Matter 
{\bf 5}, 3565 (1993);
H. R\"oder {\it et al.},
Europhys. Lett. {\bf 28}, 257 (1994);
H. Fehske {\it et al.},
Phys. Rev. B {\bf 51},  16582 (1995);
G. Wellein {\it et al.},
Phys. Rev. B {\bf 53}, 9666 (1996);
G. Wellein {\it et al.},
Physica C {\bf 282--287}, 1827 (1997);
B. B\"auml {\it et al.},
\prb {\bf 58}, 3663 (1998);
and references therein.
\item{[\olRFCCe]}
%
J. K. Freericks and M. Jarrell,
Phys. Rev. Lett. {\bf 75}, 2570 (1995);
A. Greco and A. Dobry, Solid State Commun. {\bf 99}, 473 (1996); 
D. Poilblanc {\it et al.}, Europhys. Lett. {\bf 34}, 367 (1996);
R. Fehrenbacher, 
Phys. Rev. Lett. {\bf 77}, 2288 (1996);
M. Capone {\it et al.},
\prb {\bf 56}, 4484 (1997);
%
W. Stephan {\it et al.},
Phys. Lett. A {\bf 227}, 120 (1997);
%
S. Ishihara {\it et al.},
Phys. Rev. B {\bf 55}, 3163 (1997);
T. Sakai {\it et al.},
Phys. Rev. B {\bf 55}, 8445 (1997).
%
%
\item{[\olRFDD]}
For recent reviews of the experiments, see
{\it Lattice Effects in High-$T_c$ Superconductors},
Y. Bar-Yam, T. Egami, J. Mustre-de Leon, and A. R. Bishop (eds.)
(World Scientific, Singapore, 1992); 
S.J.L. Billinge {\it et al.\/}, in {\it Strongly Correlated 
Electronic Materials: The Los Alamos Symposium 1993\/}, 
K. Bedell {\it et al.\/} (eds.) (Addison-Wesley Reading, Massachusetts, 1994); 
T. Egami and S. J. L. Billinge, 
Prog. Mater. Sci. {\bf 38}, 359 (1994);
T. Egami and S. J. L. Billinge, 
in {\it Physical Properties of High Temperature Superconductors V},
D. M. Ginsberg (ed.) (World Scientific, Singapore, 1996);
and references therein.
%
%
%
\item{[\olRFEE]} 
      C.G. Olson et al.,
        Phys. Rev. B {\bf{42}}, 381 (1990).
\item{[\olRFEEu]} 
      R. Manzke et al., 
        Physica Scripta {\bf{41}}, 579 (1990).
\item{[\olRFEEv]} 
      B.O. Wells et al.,
        Phys. Rev. Lett. {\bf{65}}, 3056 (1990).
\item{[\olRFEEw]} 
      L.Z. Liu et al., 
        J. Phys. Chem. Solids {\bf{52}}, 1473 (1991).
\item{[\olRFEEa]} 
      R. Claessen et al.,
        Phys. Rev. Lett. {\bf{69}}, 808 (1992).
\item{[\olRFEExa]} 
      D.M. King et al.,
        \prl {\bf{70}}, 3159 (1993).
\item{[\olRFEEya]} 
      R.O. King et al.,
        \prl {\bf 70}, 3163 (1993).
\item{[\olRFEEb]} 
      D.S. Dessau et al.,
        \prl {\bf{71}},  2781 (1993).
%
\item{[\olRFFF]} 
   D.J. Scalapino et al., 
     Phys. Rev. B {\bf{34}}, 8190 (1986);
     {\it ibid.} {\bf{35}}, 6694 (1987).
\item{[\olRFFFxa]} 
   K. Miyake {\it et al.}, 
     \prb {\bf{34}}, 6554 (1986).
\item{[\olRFFFya]}
N. E. Bickers {\it et al.},
Int. J. Mod. Phys. B {\bf 1}, 687 (1987);
N. E. Bickers {\it et al.},
\prl {\bf 62}, {961} (1989);
N. E. Bickers and S. R. White,
\prb {\bf 43}, {8044} (1991);
S. R. White {\it et al.},
R. T. Scalettar \prb {\bf 39}, 839 (1989).
\item{[\olRFFFza]} 
K. Yonemitsu, J. Phys. Soc. Jpn. {\bf 58}, 4576 (1989).
\item{[\olRFFFa]} 
A. Kampf and J.R. Schrieffer,
     \prb {\bf{42}}, 7967 (1990);
     {\it{ibid.}} {\bf{41}}, 6399 (1990).
\item{[\olRFFFb]} 
   T. Moriya et al., 
     J. Phys. Soc. Jap. {\bf{59}}, 2905 (1990).
   K. Ueda et al., 
   in {\it{Electronic Properties and Mechanisms
     of High-$T_c$ Superconductors}}, T. Oguchi et al. (Ed.),
     pp. 145 (North-Holland, 1992).
\item{[\olRFFFc]} 
A. Millis {\it et al.},
Phys. Rev. B {\bf 42}, 167 (1990);
H. Monien {\it et al.},
{\it{ibid.}} {\bf{43}}, 258 (1991);
H. Monien {\it et al.},
{\it{ibid.}} {\bf{43}}, 275 (1991).
P. Monthoux et al., 
     Phys. Rev. Lett. {\bf{67}}, 3448 (1991);
     Phys. Rev. B {\bf{46}}, 14803 (1992);
P. Monthoux and D. Pines,
     Phys. Rev. Lett. {\bf{69}}, 961 (1992);
     Phys. Rev. B {\bf{47}}, 6069 (1993).
\item{[\olRFFFd]} 
R.J. Radtke et al.,
   \prb {\bf{46}}, 11975 (1992);
   {\it{ibid.}} {\bf{48}}, 653 (1993).
\item{[\olRFFFg]} 
J.W. Serene and D.W. Hess, Phys. Rev. B {\bf 44}, 3391
(1991); J. Phys. Chem. Solids {\bf 52}, 1385 (1991);
P. Monthoux and D.J. Scalapino, Phys. Rev. Lett. {\bf 72}, 1874 (1994);
T. Dahm and L. Tewordt, {\it ibid.} {\bf 74}, 793 (1995);
St. Lenck, J.P. Carbotte, and R.C. Dynes, Phys. Rev. 
B {\bf 50}, 10149 (1994).
\item{[\olRFFFh]} 
John Luo and N.E. Bickers, 
Phys. Rev. B {\bf 47}, 12153 (1993); 
{\it ibid.} {\bf 48}, 15983 (1993);
C.--H. Pao and N.E. Bickers, Phys. Rev. B {\bf 49}, 
1586 (1994);
C.--H. Pao and N.E. Bickers, Phys. Rev. Lett. {\bf 72},
1870 (1994); 
C.--H. Pao and N.E. Bickers, Phys. Rev. B {\bf 49}, 
1586 (1994);
G. Esirgen and N. E. Bickers, {\it ibid.} {\bf 55},
2122 (1997); {\it ibid}. {\bf 57}, 5376 (1998);
G. Esirgen, H.--B. Sch\"uttler, 
and N. E. Bickers, cond--mat/9806264.
%
%
\item{[\olRFGGa]} 
   D.J. Scalapino, 
     in {\it{Superconductivity}}, 
       Ed. R.D. Parks, (Marcel Dekker, New York, 1969).
\item{[\olRFGGb]} 
   P. B. Allen and B. Mitrovic,
     in {\it{Solid State Physics}},
     Eds. H.Ehrenreich, F. Seitz and D. Turnbull
     (Academic Press, New York, 1982), Vol 37, pp.1-91;
   P.B. Allen,
     Comments Cond. Matt. Phys. {\bf{15}}, 327 (1992).
%
\item{[\olRFHH]} 
   T.D. Holstein, 
     Ann. Phys. (N.Y.) {\bf{8}}, 325 (1959).
\item{[\olRFHHa]} 
    T. D. Holstein,
     Ann. Phys. (N.Y.) {\bf{8}}, 343 (1959).
\item{[\olRFHHb]} 
    T. D. Holstein,
    Mol. Cryst. Liq. Cryst. {\bf{79}}, 235 (1981).
\item{[\olRFHHxb]} 
    D. Emin and T.D. Holstein,
      Phys. Rev. Lett. {\bf{36}}, 323 (1976).
\item{[\olRFHHc]} 
     H.-B. Sch\"uttler and T.D. Holstein,
      \prl, {\bf{51}} (1983).
\item{[\olRFHHd]} 
     H.-B. Sch\"uttler and T.D. Holstein,
     Ann. Phys. (N.Y.) {\bf{166}}, 93 (1986).
\item{[\olRFHHub]} 
D. Emin and T. Holstein, Ann. Phys. (N.Y.) {\bf 53}, 439 (1969).
%
\item{[\olRFIIa]} 
  J.C.K. Hui and P.B. Allen,
    J. Phys. F {\bf 4}, L42 (1974).
\item{[\olRFIIb]} 
  N.M. Plakida {\it et al.}, 
    Europhys. Lett. {\bf 4}, 1309 (1987);
T. Galbataar {\it et al.},
    Physica C {\bf 176}, 496 (1991).
\item{[\olRFIIc]} 
  J.R. Hardy and J.W. Flocken,
    Phys. Rev. Lett. {\bf 60}, 2191 (1988).
\item{[\olRFIId]} 
  V.H. Crespi et al.,
    Phys. Rev. B {\bf 43}, 12921 (1991);
  V.H. Crespi and M.L. Cohen,
    {\it{ibid.}} {\bf 44}, 4712 (1991).
\item{[\olRFIIe]} 
A. Bussmann-Holder et al.,
    Phys. Rev. Lett. {\bf 67}, 512 (1991);
    Phys. Rev. B {\bf 43}, 13728 (1991).
%
\item{[\olRFJJ]} 
    G. Mahan, 
      {\it{Many-Particle Physics}} (Plenum Press, New York, 1981).
%
\item{[\olRFSFVertCorr]} 
%
J. Hertz and K. Levin, Solid State Comm. {\bf 18}, 803 (1976);
B. Schuh and L. J. Sham , 
J. Low Temp. Phys. {\bf 50}, 391 (1983);
M. Grabowski and L. J. Sham,
Phys. Rev. B {\bf 29}, 6132 (1984); 
{\it ibid.} {\bf 37}, 3726 (1988);
A. J. Millis, Phys. Rev. B {\bf 45}, 13047 (1992);
N. Bulut, D. J. Scalapino, and S. R. White, Phys. Rev. B {\bf 47}, 2742 (1993);
J. R. Schrieffer, J. Low Temp. Phys. {\bf 99}, 397 (1995);
%
M. H. S. Amin and P. C. E. Stamp
Phys. Rev. Lett. {\bf 77}, 3017 (1996);
%
P. Monthoux,
\prb {\bf 55}, 15261 (1997);
%
A. V. Chubukov {\it et al.},
\prb {\bf 56}, 7789 (1997).
\item{[\olRFKK]} 
   G. Lehmann et al.,
     Phys. Status Solidi {\bf{37}}, K27 (1970);
   O. Jepsen and O.K. Anderson,
     Solid State Comm. {\bf{9}}, 1763 (1971);
   J. Rath and A.J. Freeman,
     Phys. Rev. B {\bf{11}}, 2109 (1975).
%
\item{[\olRFIsoExpZ]} 
G.-m. Zhao and D. E. Morris, Phys. Rev. B {\bf 51}, 16 487 (1995);
G.-m. Zhao and D. E. Morris
     Phys. Rev. B {\bf 54}, 15545 (1996);
G.-m. Zhao {\it et al.}, Phys. Rev. B {\bf 54}, 14 982 (1996);
G.-m. Zhao {\it et al.}, Phys. Rev. B {\bf 54}, 14 956 (1996).
\item{[\olRFIsoCritZ]} 
V. Z. Kresin, S. A. Wolf,
Phys. Rev. B {\bf 54}, 15543 (1996).
\item{[\olRFIsoExpTcA]} 
%
M.K. Crawford {\it et al.}, \prb {\bf 41}, 282 (1990); 
Science {\bf 250}, 1390 (1990).
%
\item{[\olRFIsoExpTcB]} 
J.P. Franck {\it et al.}, Physica C {\bf 185-189}, 1379 (1991); 
\prl {\bf 71}, 283 (1993); 
J. P. Franck and D. D. Lawrie, 
Physica C {\bf 235-240}, 1503 (1994);
%
J. Supercond. {\bf 8}, 591 (1995);
J. P. Franck, Physica Scripta {\bf T66}, 220 (1996);
%
J. P. Franck,
in {\it Physical Properties of High Temperature Superconductors IV},
D. M. Ginsberg (ed.) (World Scientific, Singapore, 1996), p. 189;
and references therein.
\item{[\olRFPaScIsoD]} 
C.-H. Pao and H.-B. Sch\"uttler, J. Supercond. {\bf 8}, 633 (1995);
C.-H. Pao and H.-B. Sch\"uttler, 
J. Phys. Chem. Solids {\bf 56},1745 (1995);
H.-B. Sch\"uttler and C.-H. Pao, 
Phys. Rev. Lett. {\bf 75}, 4504 (1995);
C.-H. Pao and H.-B. Sch\"uttler, 
\prb {\bf 57}, {5051} (1998);
C.-H. Pao and H.-B. Sch\"uttler,
{}cond-mat/9809321.
\item{[\olRFAlleTr]} 
P. B. Allen, Phys. Rev. B {\bf 3}, 305 (1971);
{\it{ibid.}} {\bf 17}, 3725 (1978);
Comments Cond. Matter Physics {\bf 15}, 327 (1992);
P. B. Allen, W. E. Pickett, and H. Krakauer,
Phys. Rev. B {\bf 37}, 7482 (1988).
\item{[\olRFDruOpt]} 
For reviews of optical conductivity measurements in the cuprates, see
T. Timusk and D.B. Tanner,
in {\it Physical Properties of High Temperature Superconductors I}, 
D. M. Ginsberg, (ed.) (World Scientific, 1989);
M. Suzuki, in {\it{Strong Correlation and Superconductivity}},
H. Fukuyama, S. Maekawa, and A. Malozemoff (eds.)
(Springer, Berlin, 1989);
T. Timusk and D.B. Tanner, 
in {\it Physical Properties of High Temperature Superconductors III}, 
D. M. Ginsberg, (ed.) (World Scientific, 1992);
and references therein.
\item{[\olRFEmBP]} 
D. Emin, \prl {\bf 62}, 1544 (1989);
D. Emin and M. S. Hillery, \prb {\bf 39}, 6575 (1989);
D. Emin, {\it ibid.} {\bf 48}, 13691 (1993);
{\it ibid.} {\bf 49}, 9157 (1994);
\prl {\bf 72}, 1052 (1994).
\item{[\olRFLL]} In Ref.~[\olRFFFa], the $\rho(\omega)$-resonance for the pure SF
model was not found, apparently due to lack of
resolution in the numerical ${\vec{k}}$-integration. 
(A. Kampf, private communication).
\item{[\olRFMM]} A similar $\rho(\omega)$-peak is discussed
for the infinite-dimensional Hubbard model by
   M. Jarrell and T. Pruschke 
     Z. Phys. B {\bf{90}}, 187 (1993);
   Th. Pruschke, {\it et al.},
     Phys. Rev. B {\bf 47}, 3553 (1993);
   X.Y. Zhang et al.,
      Phys. Rev. Lett. {\bf{70}}, 1666 (1993);
and for the 2D Hubbard model by
   N. Bulut {\it et al.},
      Phys. Rev. Lett. {\bf{72}}, 705 (1994).
%
%
\item{[\olRFNN]} 
To what extent the measured ARPES line shapes 
really reflect the single-particle
spectrum $P({\vec{k}},\omega)$ of a quasi-2D material
is still controversial; see Ref.~[\olRFEEa] and N.V. Smith,
Comments Cond. Matt. Phys. {\bf{15}}, 263 (1992).
%
\item{[\olRFOO]} 
   R.J. Radtke et al.,
     Phys. Rev. B {\bf{48}}, 15957 (1993);
R. J. Radtke and M. R. Norman,
     {\it ibid.} {\bf 50}, 9554 (1994).
%
\item{[\olRFPP]} D.M. Newns et al.,
Comments Cond. Matt. Phys. {\bf{15}}, 273 (1992).
\hbn
\vfil
\eject
\centerline{FIGURE CAPTIONS}

\hbn
FIG.~1.
Mass enhancement 
$Z\!=\!1-\partial_\omega\Sigma^\prime({\vec{k}}_F,\omega=0)$ 
vs. temperature $T$ at the Fermi
surface point ${\vec{k}}_F(1,1)\!\cong\!0.44(\pi,\pi)$
in the APSF model, with $V\!\equiv\!V_{AP}+V_{SF}$ in Eq.~(25),
(upper full line)
and in the purely electronic SF model, with $V\!\equiv\!V_{SF}$ in Eq.~(25)
(lower full line).
Also shown is $Z_{AP}\!=\!-\partial_\omega\Sigma^\prime_{AP}(\omega=0)$ 
vs. T in the APSF model (dashed line), where $\Sigma^\prime_{AP}$ is the 
$V_{AP}$-contribution to  $\Sigma^\prime$ in the APSF model.
The model parameters are given in the text in Sec.III.A, following Eq.~(46).
\hbn
\hbn
FIG.~2. 
Density of states $\rho(\omega)$ vs. energy $\omega$ 
at temperature $T\!\cong\!\Omega_t$.
APSF (full line) and SF (dashed line) are the 
results, for the respective models, as defined in Fig.~1.
The results labeled SF1 (dot-dash line) are  results
{}of non-self-consistent 1{}st order calculations
for the SF model, as descibed in text.
Inset: APSF and SF results for $\rho(\omega)$ vs. $\omega$
near $\omega\!=\!0$ at $T\!\cong\!\Omega_t$ and $T\!=\!0$.
The other model parameters are the same as used in FIG.~1.
\hbn
\hbn
FIG.~3.
Hole excitation spectra
$P({\vec{k}},\omega)$ vs. energy $\omega$ in the APSF model
[with $V\!\equiv\!V_{AP}+V_{SF}$ in Eq.~(25)] for 
${\vec{k}}\!=\!(k_x,k_x)$ at $k_x$-values indicated,
at temperature $T\!\cong\!\Omega_t$.
Spectra are off-set by successive vertical shifts,
with dashed lines showing zero intensity.
The other model parameters are the same as used in FIG.~1.
\hbn
\hbn
FIG.~4.
Quasi-particle bands
$E_{\vec{k}}$ vs. wavevector ${\vec{k}}$ in the APSF model 
[with $V\!\equiv\!V_{AP}+V_{SF}$ in Eq.~(25), full line]
and in the non-interacting system 
[$V\!\equiv\!0$, dashed line], along several directions in the
2D Brillouin zone, both
at the same electron concentration $\langle n_j\rangle\!=\!0.75$, 
at temperature $T\!\cong\!\Omega_t$.
The other model parameters are the same as used in FIG.~1.
\hbn
\vfil
\eject
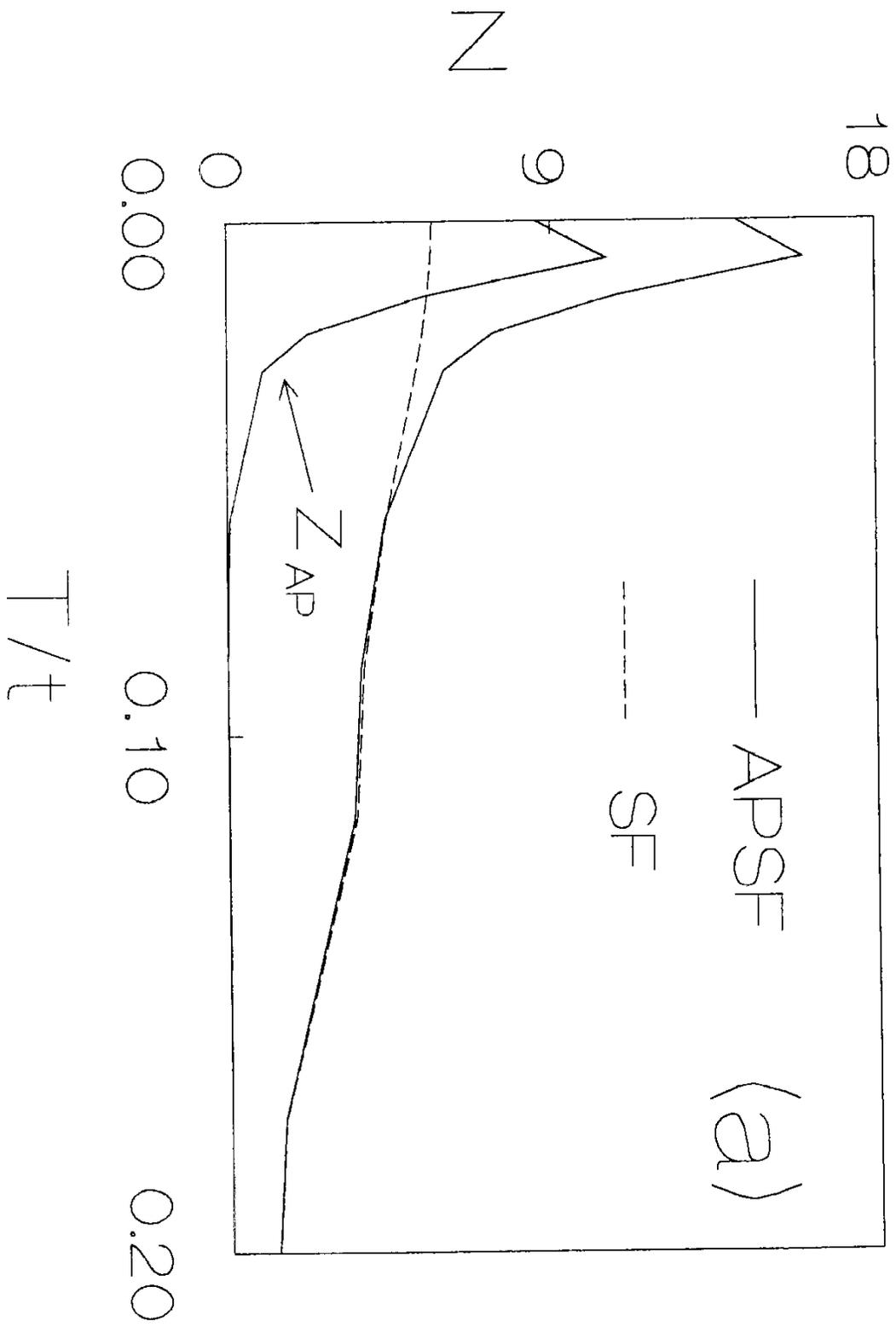
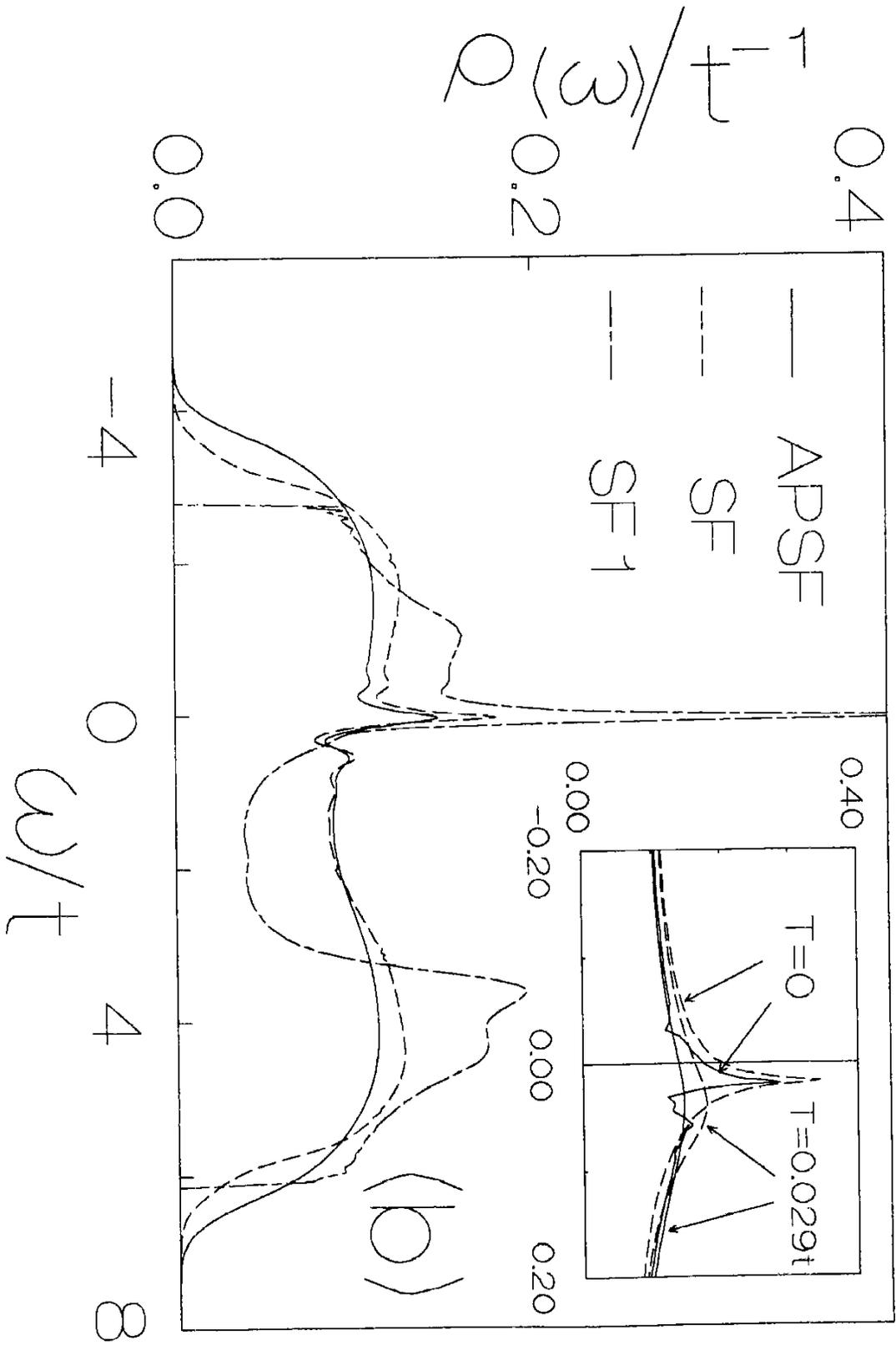
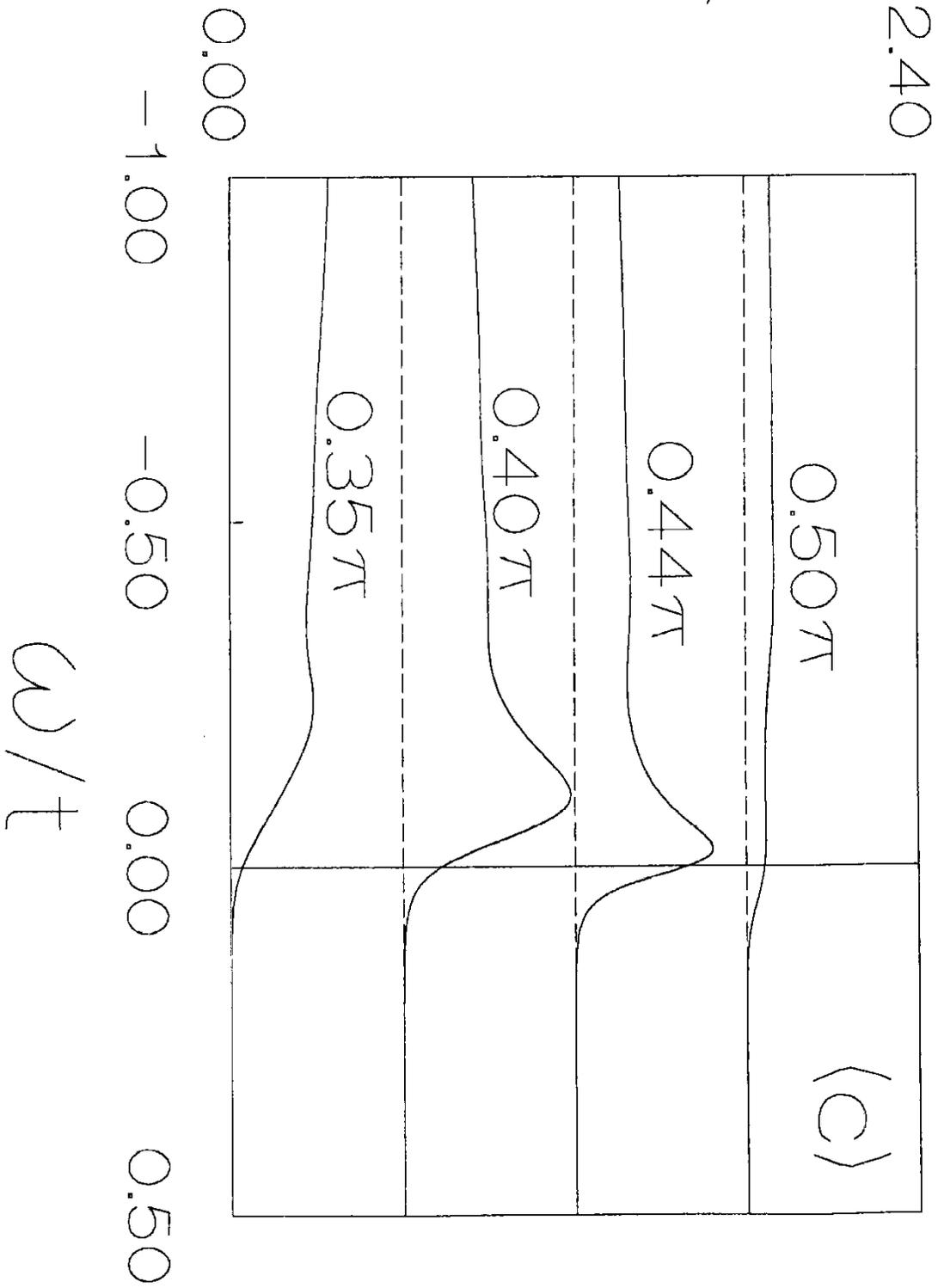
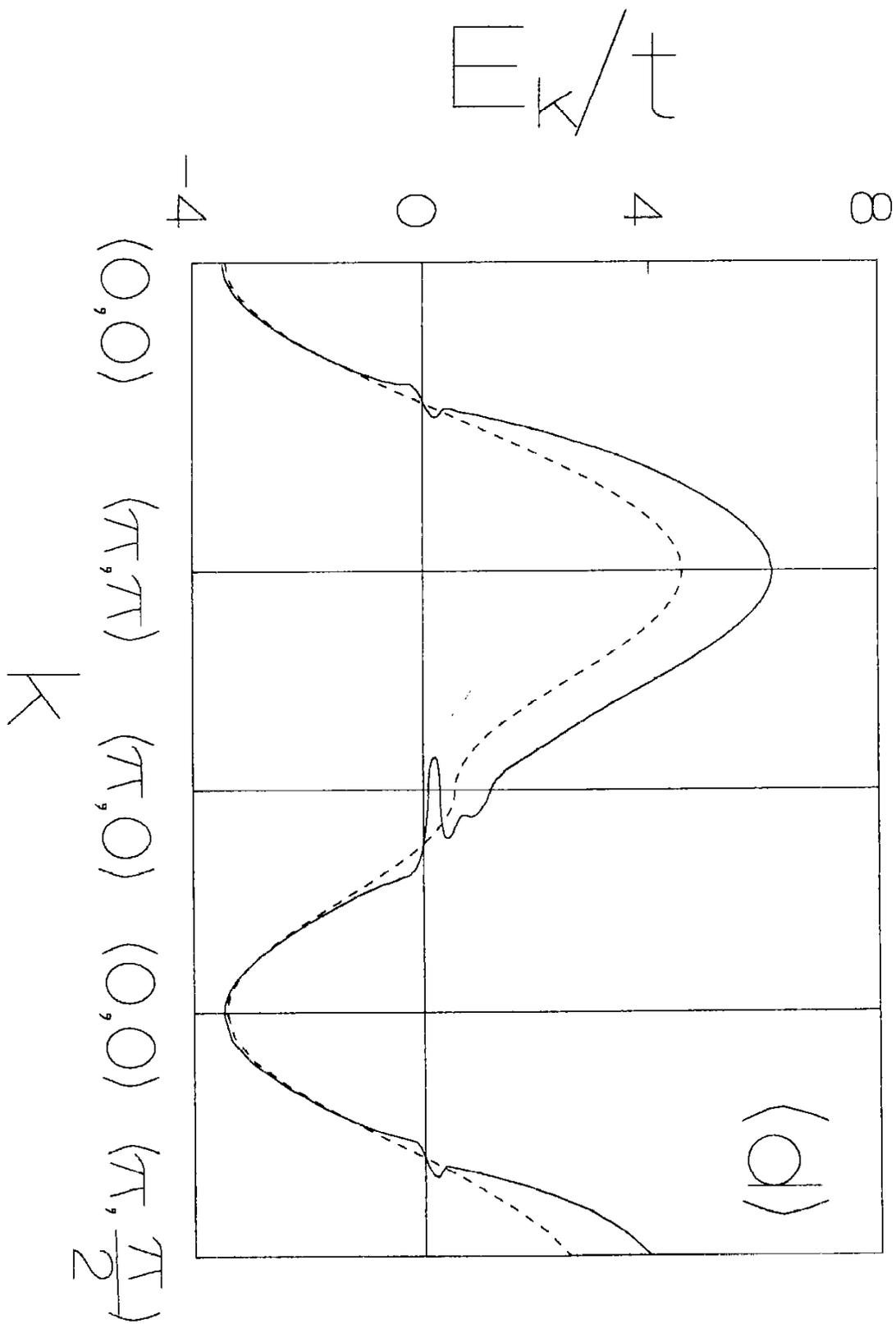
\bye